\documentclass[aps,prx,a4,superscriptaddress,preprintnumbers,twoside,twocolumn,floatfix]{revtex4}
\usepackage{stmaryrd}
\usepackage{graphicx}
\usepackage{color}
\usepackage{bm,bbold}
\usepackage{bbm}
\usepackage{enumerate}
\usepackage{color}
\graphicspath{figures} % Location of the graphics files
\usepackage{amsfonts, amsmath, amsthm, amssymb} 
\usepackage{mathtools}
\usepackage{dsfont}
\usepackage{array}
\usepackage{changes}

\usepackage[colorlinks,bookmarks=false,citecolor=blue,linkcolor=red,urlcolor=blue]{hyperref}

\begin{document}

\title{Inferring non-linear many-body Bell's inequalities from average two-body correlations:\\
Systematic approach for arbitrary spin-$j$ ensembles}

\author{Guillem M\"uller-Rigat}
\affiliation{ICFO - Institut de Ciencies Fotoniques, The Barcelona Institute of Science and Technology, 08860 Castelldefels (Barcelona), Spain}

\author{Albert Aloy}
\affiliation{ICFO - Institut de Ciencies Fotoniques, The Barcelona Institute of Science and Technology, 08860 Castelldefels (Barcelona), Spain}

\author{Maciej Lewenstein}
\affiliation{ICFO - Institut de Ciencies Fotoniques, The Barcelona Institute of Science and Technology, 08860 Castelldefels (Barcelona), Spain}
\affiliation{ICREA, Pg. Llu\'{\i}s Companys 23, 08010 Barcelona, Spain}

\author{Ir\'en\'ee Fr\'erot}
\email[]{irenee.frerot@gmail.com}
\affiliation{ICFO - Institut de Ciencies Fotoniques, The Barcelona Institute of Science and Technology, 08860 Castelldefels (Barcelona), Spain}
\affiliation{Max-Planck-Institut f{\"u}r Quantenoptik, D-85748 Garching, Germany}

\date{\today}

\begin{abstract}
Violating Bell's inequalities (BIs) allows one to certify the preparation of entangled states from minimal assumptions -- in a device-independent manner. Finding BIs tailored to many-body correlations as prepared in present-day quantum computers and simulators is however a highly challenging endeavour. {In this work, we focus on BIs violated by very coarse-grain features of the system: two-body correlations averaged over all permutations of the parties. For two-outcomes measurements, specific BIs of this form have been theoretically and experimentally studied in the past, but it is practically impossible to explicitly test all such BIs. Data-driven methods -- reconstructing a violated BI from the data themselves -- have therefore been considered. Here, inspired by statistical physics, we develop a novel data-driven approach specifically tailored to such coarse-grain data. Our approach offers two main improvements over the existing literature: 1) it is directly designed for any number of outcomes and settings; 2) the obtained BIs are quadratic in the data, offering a fundamental scaling advantage for the precision required in experiments. This very flexible method, whose complexity does not scale with the system size, allows us to systematically improve over all previously-known Bell's inequalities robustly violated by ensembles of quantum spin-$1/2$; and to discover novel families of Bell's inequalities, tailored to spin-squeezed states and many-body spin singlets of arbitrary spin-$j$ ensembles.}
\end{abstract}

\maketitle

\section{Introduction}
Multipartite entanglement is a central feature of quantum many-body systems, fundamentally challenging our ability to efficiently simulate them on classical computers \cite{georgescuetal2014,deutsch2020}. For the same reason, quantum entanglement distributed among many degrees of freedom represents a key resource for quantum simulators and computers. 
Consequently, proving that the multipartite states prepared in quantum simulators or computers are indeed entangled -- namely, the task of entanglement certification -- is a key step in assessing the quantum advantage offered by such devices. Depending on the assumptions made about the individual components of the device, two different paradigms may appear suitable. In a so-called device-dependent framework, the subsystems are well characterized: the Hilbert space is known (e.g. a qubit space), and the measurements correspond to well-defined quantum observables (e.g. spin measurements); in this framework, entanglement certification relies on the violation of a certain entanglement witness \cite{guhne_entanglement_2009}. On the other hand, in a device-independent framework, no assumption is made about the Hilbert space of the subsystems, and consequently the measurements correspond to unknown quantum observables; this framework appears especially suitable when considering effective few-level systems, where the actual Hilbert space can contain an unlimited number of physical degrees of freedom. Relaxing certain assumptions about the system clearly makes entanglement certification more demanding; nevertheless, device-independent entanglement certification is possible if the violation of a certain Bell's inequality \cite{brunneretal2014} can be established. {Designing many-body Bell tests is the focus of the present paper.

\begin{figure*}[t]
	\includegraphics[width=0.9\textwidth]{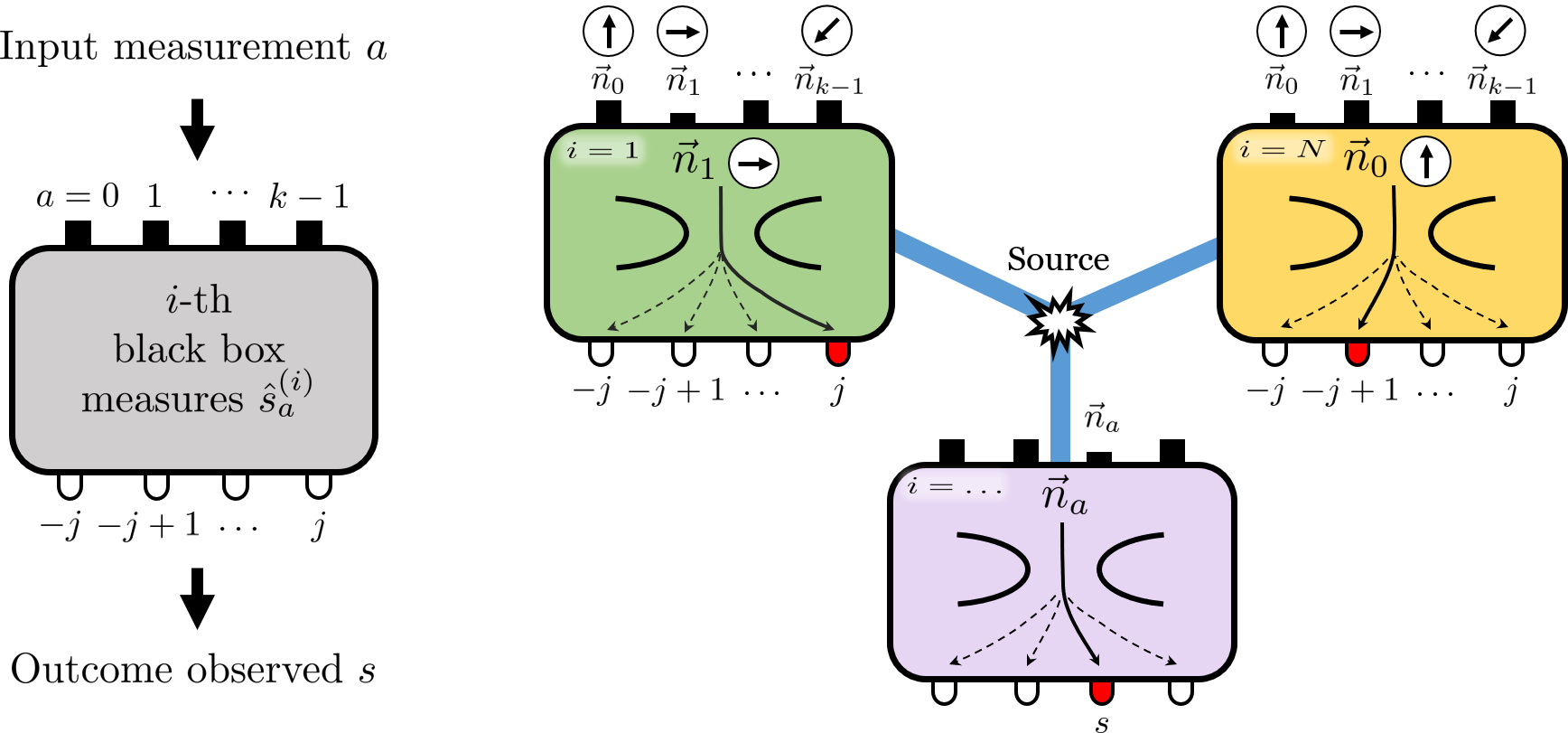}
\caption{Illustration of an ideal multipartite Bell test for entanglement certification. Right: A composite quantum system prepared by the source is shared among $N$ spatially-separated observers (on the sketch, $N=3$), each of which chooses a measurement settings $a\in\{0,1,\ldots,k-1\}$, obtaining an outcome $s\in\{-j,-j+1,\ldots,j\}$. In this work, we focus on spin measurements on quantum spin-$j$ particles, sketched as Stern-Gerlach magnets oriented along directions $\vec{n}_a$ -- but Bell tests are independent of these assumptions. This procedure is repeated several times in order to accumulate statistics, where at each round the measurement settings and the observed outcomes may vary. If the observed statistics exhibit Bell's non-local correlations, one certifies that the multipartite state is quantum-mechanically entangled. Left: In a Bell test, each subsystem $i$ is treated as a black box, with no assumption about the Hilbert space. The actual observables $\hat{s}_a^{(i)}$ being measured are attributed arbitrary input labels $a$, and the measurement outcomes $s$ have no physical meaning. Entanglement is therefore certified in a device-independent manner.}
	\label{BellExperimentSpin}	
\end{figure*}

 As fully characterizing the many-body correlations among the subsystems requires exponentially-many measurements, any scalable Bell test must rely on incomplete information, obtained from an accessible number of measurements -- for instance, the knowledge of few-body correlations among the parties \cite{baccarietal2017,wangetal2017,frerotR2020}. In particular, in order to mitigate scalability issues, a successful strategy is to symmetrize the data from which Bell non-locality is to be certified, over all permutations of the parties \cite{turaetal2014,schmiedetal2016,engelsen_bell_2017,fadelT2017,wagner_bell_2017,frerotR2020}. Correspondingly, the Bell's inequalities relevant to such coarse-grain features of the system involve a number of coefficients which is independent of the number of parties. On the experimental side \cite{schmiedetal2016,engelsen_bell_2017}, this allows one to reduce the statistical uncertainty on the data entering the Bell's inequality; on the theoretical side, this allows one to reduce the computational complexity, often leading to the analytical characterization of Bell's inequalities \cite{turaetal2014,wagner_bell_2017,frerotR2020}. A second challenge for entanglement certification is to take advantage, to the largest possible extent, of all the available information, without \textit{a priori} knowing the structure of entanglement within the many-body state. In particular, failing to violate all known Bell's inequalities does not imply that device-independent entanglement certification is impossible based on the available data. This motivates the development of \textit{data-driven} methods \cite{baccarietal2017,fadelT2017,frerotR2020}, where the data serve as input into an algorithm which builds, from the data themselves, a tailored Bell's inequality. In the case of two-outcome measurements -- especially suited to spin-$1/2$ systems -- a data-driven method for permutationally-invariant Bell's inequalities based on semi-definite-positive relaxations, has been proposed in the past \cite{fadelT2017}. To our knowledge, this method has however not lead to the discovery of new families of Bell's inequalities; and its extension to more outcomes -- relevant to spin-$j>1/2$ systems -- has never been achieved.}

{Taking inspiration from statistical physics, in this work we develop an alternative flexible data-driven method which takes, as input data, one- and two-body correlation functions averaged over all permutations of the subsystems -- for any number of measurement outcomes and settings. Similarly to the method of ref.~\cite{fadelT2017}, the complexity of our algorithm is independent of the system size, and tests exhaustively an infinite number of Bell's inequalities in a data-driven fashion. This lead us to recover tigher versions of all previously known permutationally-invariant Bell's inequalities \cite{turaetal2014,wagner_bell_2017,frerotR2020} in an unbiased way. Furthermore, in contrast to ref.~\cite{fadelT2017}, our scheme is directly applicable to any number of outcomes, and is validated by the study of quantum spin $j > 1/2$ ensembles. Finally, the Bell's inequalities inferred by our method are non-linear in the input data, and tightly wrap around the polytope of local-variable models (see Fig.~\ref{fig_tilde_vs_nontilde} for an example). This feature offers a fundamental scaling improvement regarding experimental requirements, including for all previously-known Bell's inequalities invariant under permutations \cite{turaetal2014,schmiedetal2016,engelsen_bell_2017,wagner_bell_2017,frerotR2020}.
 Among other results obtained with our novel method, we discover new families of many-body Bell's inequalities, for measurements involving arbitrarily-many outcomes, violated by paradigmatic many-body entangled states for ensembles of quantum spins $j\ge 1/2$ -- namely, spin singlets and spin-squeezed states -- a topic of timely experimental revelance to many experimental platforms manipulating qudit ensembles. 
 
Before entering into the details of our new method, in the rest of this section we review the framework of device-independent entanglement certification (\ref{sec_DIEC}), and the notion of Bell's inequalities invariant under permutations (\ref{sec_PIBI_concept}). In Section \ref{sec_d2}, we present our method in the case of two-outcomes measurements (\ref{sec_convex_opt_algo_d2}), and apply it to improve over and extend previously-known Bell's inequalities in the case of spin-singlets (\ref{sec_BI_singlets_d2}) and spin-squeezed states (\ref{sec_PIBI_squeezed}) for spin-$1/2$ ensembles. In particular, we emphasize the fundamental scaling improvement offered by the non-linear nature of the Bell's inequalities inferred by our algorithm (see e.g.~Fig.~\ref{fig_tilde_vs_nontilde}). Section \ref{sec_many_outcomes} is then devoted to the hitherto-unexplored case of spin-$j>1/2$ (namely, qudits) ensembles, for which we extend our approach to an arbitrary number of outcomes (\ref{sec_algo_spin_j}). We then apply it again to spin-singlets (\ref{sec_singlets_spin_j}) and spin-squeezed state (\ref{sec_squeezing_spin_j}), leading us to characterize analytically novel families of Bell's inequalities, valid for any number of parties and outcomes. Section \ref{sec_experimental} contains experimental considerations: in Section \ref{sec_exp_platforms}, we list different platforms and their respective capabilities to detect entanglement and Bell correlations; in Section \ref{sec_experimental_noise} we discuss the statistical requirements to acquire the data used as input to our algorithm. Finally, Section \ref{sec_conclusions} displays our conclusions and prospects.
}

\subsection{Device-independent entanglement certification} 
\label{sec_DIEC}
\textbf{Bell scenario.}
Arguably, the violation of Bell's inequalities \cite{brunneretal2014} represents the most robust scheme to certify entanglement, avoiding detailed assumptions about the physical nature of the degrees of freedom being measured, and about the accurate calibration of the measurements being performed. In this so-called device-independent scenario (Fig.~\ref{BellExperimentSpin}), each subsystem $i \in \{1, \dots N\}$ (for instance, a quantum spin-$j$) is treated as a black box, namely, no assumption is made about the actual Hilbert space of the system. This black-box treatment is especially relevant when dealing with effective few-level systems. On each subsystem, $k$ different measurement settings can be implemented. In practice, they correspond to certain quantum operators $\hat s_a^{(i)}$ ($a \in \{0, \dots, k-1\}$), for instance spin measurements along particular directions ${\bf n}_a$, but in a device-independent scenario the actual quantum measurement which is performed is not assumed; instead, only the outcome of the measurement, denoted $s$, is collected (see Fig.~\ref{BellExperimentSpin}) {(throughout the paper, we denote as $\hat{\cal O}$ a quantum observable, and as ${\cal O}$ the outcome of its measurement)}. The only assumption made is that the number $d$ of possible outcomes for each measurement is finite. In practice, the possible values of $s$ are the eigenvalues of the quantum operator $\hat s_a^{(i)}$ (for instance, the $2j+1$ possible values of a spin-$j$ measurement), but in a device-independent scenario these are mere labels for the outcomes, with no specific physical meaning. For convenience and later connection with quantum violations of Bell's inequalities when performing appropriate spin measurements on quantum many-body systems, we denote these possible outcomes as $s \in \{-j, -j+1, \dots, j\}$ with $d=2j+1$ -- but it should be emphasized that within a device-independent framework, these labels are arbitrary. A Bell experiment \cite{brunneretal2014} consists in repeating the following sequence: 1) choose a setting $a^{(i)} \in \{0, \dots, k-1\}$ for each subsystem; 2) perform the corresponding measurements, yielding the $N$ outcomes ${\bf s}= \{s^{(i)}\}_{i=1}^N$. By repeating this sequence, varying the measurement settings ${\bf a} = \{a^{(i)}\}_{i=1}^N$, statistics of the measurement outcomes are collected. Complete information is obtained if one reconstructs all $N$-body marginal probability distributions: $P({\bf s}|{\bf a})$ for all choices of settings ${\bf a}$. If one denotes $\hat\Pi_{a,s}^{(i)}$ the projector onto the eigenspace of the observable $\hat s_a^{(i)}$ associated to the eigenvalue $s$, then these probabilities are given by
\begin{equation}
	P({\bf s}|{\bf a}) = {\rm Tr}[\hat \rho \otimes_{i=1}^N \hat\Pi_{a^{(i)}, s^{(i)}}^{(i)}] ~,
\end{equation}
where $\hat\rho$ is the density-matrix of the system -- notice that even if, in a device-independent scenario, we remain agnostic about the Hilbert space over which $\hat\rho$ acts, such a decomposition exists in principle.\\

\textbf{Bell's inequalities and entanglement certification.} The state $\rho$ is not entangled (namely, it is separable) if it can be decomposed as a statistical mixture of product states:
\begin{equation}
	\hat\rho_{\rm sep} = \int_{\lambda} d\mu(\lambda) \otimes_{i=1}^N \hat\rho_\lambda^{(i)} ~,
	\label{eq_rho_sep}
\end{equation} 
where $\hat\rho_\lambda^{(i)}$ is an arbitrary local quantum state (pure or mixed) for subsystem $i$, acting on the local Hilbert space whose dimension is arbitrary. $\lambda$ is some classical random variable, sampled with probability measure $d\mu(\lambda)$, which encodes classical correlations among the local quantum states $\hat \rho_\lambda^{(i)}$. 
The central observation behind device-independent entanglement certification is that if $\rho$ is separable [Eq.~\eqref{eq_rho_sep}], then $P({\bf s}|{\bf a})$ can always be reproduced by a local-variable (LV) model in the sense of J.~S.~Bell \cite{bell1964,brunneretal2014}:
\begin{equation}
	P_{\rm LV}({\bf s}|{\bf a}) = \int_\lambda d\mu(\lambda)~ \prod_{i=1}^N P^{(i)}_\lambda[s^{(i)}|a^{(i)}]~,
	\label{eq_P_LV}
\end{equation}
where $P^{(i)}_\lambda(s|a) = {\rm Tr}[\hat \rho_\lambda^{(i)} \hat\Pi^{(i)}_{a,s}]$. In a device-independent framework, we do not know the explicit expressions of the projectors $\hat\Pi_{a, s}^{(i)}$ corresponding to the measurements which are actually being performed; and even the Hilbert space of the system, over which the quantum state $\hat \rho$ and these projectors are defined, is unknown and arbitrary -- it could even be infinite dimensional. Yet, regardless of the actual Hilbert space describing the system, if the state is not entangled, then a decomposition as in Eq.~\eqref{eq_P_LV} must exist for the experimentally-observed correlations contained in $P({\bf s}|{\bf a})$.

 Therefore, if conversely $P({\bf s}|{\bf a})$ is found to violate a Bell's inequality -- denying the possibility to decompose $P({\bf s}|{\bf a})$ as in Eq.~\eqref{eq_P_LV}--, then $\hat \rho$ must be entangled. Crucially, this holds regardless of the Hilbert space of the individual subsystems, and regardless of the measurements which were actually performed to generate $P({\bf s}|{\bf a})$. Violating a Bell's inequality therefore certifies that $\hat \rho$ is entangled in a device-independent manner. 
 
Note that in principle, violating a Bell's inequality allows for quantum information protocols more powerful than merely witnessing entanglement \cite{brunneretal2014}, which is the task on which we focus in this paper.

\subsection{Permutationally-invariant Bell's inequalities from two-body correlations}
\label{sec_PIBI_concept}
\textbf{Certifying entanglement from two-body correlations.}
 Overall, reconstructing $P({\bf s}|{\bf a})$ requires collecting $k^N(d^N-1)$ probabilities. This exponential scaling clearly makes full reconstruction of these marginals unpractical, and therefore, methods based on partial information have been developed. The simplest non-trivial strategy, which we follow in this paper, is to consider jointly all two-body marginals: $P^{(ij)}(s,t|a,b)$ ($i \neq j$), namely the probability to obtain the pair of outcomes $(s,t)$ if measurement $a$ is performed on subsystem $i$, and measurement $b$ on subsystem $j$, for all possible pairs of subsystems $1\le i < j \le N$, and all possible pairs of measurement settings $0 \le a,b \le k-1$.\\

\textbf{Local-variable models as distributions over classical spin configurations.}
The \textit{\`a-la-Bell} formulation of LV models as in Eq.~\eqref{eq_P_LV} makes transparent the link with entangled quantum states. It is however more intuitive to represent LV models as probability distributions over the measurement results treated as classical variables $s_a^{(i)}$ \cite{frerotR2020}. Indeed, as first proved by Fine \cite{Fine1982,brunneretal2014}, a LV decomposition as in Eq.~\eqref{eq_P_LV} exists if and only if there exists a grand-probability distribution $P_{\rm LV}[{\bm \sigma}]$ over the fictitious ensemble of classical variables {${\bm \sigma} = \{s_a^{(i)};~a=0,\dots,k-1;~i=1,\dots,N\}$, such that the observed statistical properties are obtained as marginals against $P_{\rm LV}$, i.e. \cite{Fine1982,frerotR2020}:
\begin{equation}
	P_{\rm LV}^{(ij)}(s,t|a,b) = \sum_{{\bm \sigma}\in \{-j,\dots,j\}^{kN}} P_{\rm LV}[{\bm \sigma}] ~\delta_{s_a^{(i)},s}\delta_{s_b^{(j)},t}~,
	\label{eq_Ppair_marginal_PLV}
\end{equation} }
where $\delta$ is the Kronecker symbol ($\delta_{x,y} = 1$ if $x=y$, and $\delta_{x,y}=0$ otherwise). In LV models, measurement results may therefore be viewed as sampled from an underlying classical ``spin'' configuration ${\bm \sigma}$, where $k$ ``hidden'' $d-$level spins $s_{a}^{(i)} \in \{-j, -j+1, \dots j\}$ are attached to each subsystem $i$, encoding the outcome of the measurement. While, at each measurement run with setting ${\bf a} = \{a^{(i)}\}_{i=1}^N$, the value of only one of the $k$ hidden spins is revealed [namely, $s_{a^{(i)}}^{(i)}$], in LV models all the unobserved outcomes [$s_b^{(i)}$, for $b \neq a^{(i)}$] also objectively exist independently of the act of their measurement. This contradicts standard interpretations of quantum physics if they correspond to incompatible quantum observables performed on the same subsystem, $[\hat s_a^{(i)}, \hat s_{b}^{(i)}] \neq 0$ -- and is categorically excluded if the $P^{(ij)}(s,t|a,b)$ violate a Bell's inequality, and if actions-at-a-distance are not allowed \cite{brunneretal2014}.\\

\textbf{Permutationally-invariant Bell's inequalities.}
Deciding whether or not the marginals $P^{(ij)}(s,t|a, b)$ are compatible with a grand-probability $P_{\rm LV}({\bm \sigma})$ can be mapped onto a so-called inverse Ising problem \cite{frerotR2020}, which can generically be solved in polynomial time by Monte-Carlo methods -- while worst-case instances are exponentially hard. A convergent hierarchy of relaxations to this problem has also been developed, whose computational cost is strictly polynomial at each relaxation level \cite{baccarietal2017}. Here, we drastically simplify the problem by further symmetrizing the data over all permutations of the subsystems \cite{turaetal2014,fadelT2017}, which leads us to introduce:
\begin{equation}
	\bar{P}(s,t|a, b) = \frac{1}{N(N-1)} \sum_{i \neq j} P^{(ij)}(s,t|a, b) ~.
	\label{eq_Pbar}
\end{equation} 
Bell's inequalities are constraints, of the form:
\begin{equation}
	f(\bar{P}_{\rm LV}) \ge B_{\rm c} ~
\end{equation} 
where $f$ is some function, and $B_{\rm c}$ the so-called classical bound, obeyed by all distributions $\bar{P}_{\rm LV}(s,t|a,b)$ which descend from a grand-probability $P_{\rm LV}({\bm \sigma})$. If the particular $\bar{P}$ under investigation happens to violate such a Bell's inequality [namely, if $f(\bar{P}) < B_{\rm c}$], then no grand-probability $P_{\rm LV}({\bm \sigma})$ can ever explain the data, which in turn implies that the quantum state $\hat \rho$ of the system must be entangled.

Our main result is to construct a very flexible data-driven algorithm, whose complexity is independent of $N$, allowing one to build a Bell's inequality violated by the data $\bar{P}(s,t|a,b)$ (Section \ref{sec_convex_opt_algo_d2}). This allows us to recover all previously known permutationally-invariant Bell's inequalities which are robustly violated by appropriate quantum states in the thermodynamic limit \cite{schmiedetal2016,frerotR2020}, to improve these Bell's inequalities by considering more measurement settings (Section \ref{sec_d2}), and to generalize them to scenarios with arbitrarily-many outcomes (Section \ref{sec_many_outcomes}). We discuss the potentialities of several experimental platforms to observe Bell non-locality in Section \ref{sec_experimental}, and draw our conclusions in Section \ref{sec_conclusions}.

\section{Two-outcomes measurements}
\label{sec_d2}
\textbf{Summary of the main results.}
In this section, we introduce our method by focusing on the simplest situation where the measurements can only deliver $d=2$ outcomes. The method itself is presented in Section \ref{sec_convex_opt_algo_d2}: the key results are contained in Eqs.~\eqref{eq_BI_Ct}, \eqref{eq_def_L_cost} and \eqref{eq_SDP_Ct}, which form the core of our data-driven algorithm. To be practically useful, the algorithm must be fed with carefully-chosen quantum data. In Section \ref{sec_spin_measurements}, we consider a situation where the data correspond to spin measurements on a collection of quantum spin-$1/2$. We expose the dependence of these data on collective-spin fluctuations, as represented by Eq.~\eqref{eq_Qdata_spin_half}. We then begin our data-driven exploration of Bell's inequalities with spin singlets (Section \ref{sec_BI_singlets_d2}) and spin-squeezed states (Section \ref{sec_PIBI_squeezed}), for which permutationally-invariant Bell's inequalities are already known. In both cases, we find tighter Bell's inequalities, leading to sufficient ``witness'' conditions on collective spin fluctuations which are easier to satisfy than existing ones. Concerning singlets (Section \ref{sec_BI_singlets_d2}), our main finding is a family of Bell's inequalities for arbitrarily-many measurement settings [Eq.~\eqref{eq_BI_singlets_spin_half}]. The corresponding witness condition is contained in Eq.~\eqref{eq_condition_singlet_spin_half}. Concerning spin-squeezed states (Section \ref{sec_PIBI_squeezed}), we illustrate the generic improvement offered by the non-linear nature of our Bell's inequalities on Fig.~\ref{fig_tilde_vs_nontilde}. We then go beyond existing Bell's inequalities \cite{turaetal2014,schmiedetal2016,engelsen_bell_2017} by adding an extra measurement setting, whose advantage for entanglement certification is illustrated on Figs.~\ref{fig_new_PIBI_squeezed_state} and \ref{fig_new_PIBI_squeezed_state_phase_diagram}.

\subsection{A convex-optimization algorithm}
\label{sec_convex_opt_algo_d2}
We assume that all measurements $\hat s_a^{(i)}$ can only deliver $d=2$ possible outcomes, denoted $s=\pm 1/2$ (the usual convention in the literature would be to denote them $s=\pm 1$, but we follow our general convention $s \in \{-j, -j+1, \dots j\}$ with $d=2j+1$; as already emphasized, these labels are arbitrary). Instead of working with the pair probability distribution $P^{(ij)}(s,t|a,b)$, we equivalently consider one- and two-body correlations $\langle s_a^{(i)} \rangle$ and $\langle s_a^{(i)} s_b^{(j)} \rangle$ (the two representations are related by elementary linear transformations). As coarse-grain features of the experimental data, equivalently to the averaged pair distribution $\bar{P}(s,t|a,b)$, we consider the one- and two-body correlations summed over all permutations of the subsystems:
\begin{subequations}
\label{eq_def_C}
\begin{align}
	&M_a = \sum_{i=1}^N \langle s_a^{(i)} \rangle \\
	&C_{ab} = \sum_{i \neq j} \langle s_a^{(i)} s_b^{(j)} \rangle  ~. \label{eq_def_C_1}
\end{align}
\end{subequations}
In a LV description, the $s_a^{(i)}$ are $Nk$ classical Ising spins (with values $\pm 1/2$). A LV model compatible with the (coarse-grain) experimental data corresponds to a probability distribution $P_{\rm LV}(\{s_a^{(i)}\})$ over the configurations of these Ising spins, such that $M_a$ and $C_{ab}$ are obtained as marginals against $P_{\rm LV}$. Let us assume that a LV model fitting the data exists, and derive necessary conditions obeyed by the corresponding $M_a$ and $C_{ab}$ (namely, Bell's inequalities). We first introduce the collective variables $S_a = \sum_{i=1}^N s_a^{(i)}$, and their fluctuations $\delta S_a =  S_a - \langle S_a \rangle$, so that we have:
 \begin{subequations}
 \label{eq_def_Ct}
 \begin{align}
	M_a &= \langle S_a \rangle \\
	\tilde{C}_{ab} := C_{ab} - M_a M_b &= \langle \delta S_a  \delta S_b \rangle - \sum_{i=1}^N \langle s_a^{(i)} s_b^{(i)} \rangle ~. \label{eq_def_Ct_1}
\end{align}
\end{subequations}
The terms on the r.h.s of Eq.~\eqref{eq_def_Ct_1} are not directly observable. In particular, the terms $\langle s_a^{(i)} s_b^{(i)} \rangle$ correspond to correlations among the measurement settings $a$ and $b$ on the same subsystem $i$. In the general case, these settings correspond to incompatible quantum observables, $[\hat s_a^{(i)}, \hat s_b^{(i)}] \neq 0$, and therefore these terms do not have a direct meaning in quantum physics. However, they are perfectly well-defined in LV models. The first key observation, which underlies the method developed in the present paper, is that for any $k \times k$ positive semi-definite (PSD) matrix $A \succeq 0$, and for any configuration of the collective variables $S_a$, we have $\sum_{a,b} \delta S_a A_{ab}  \delta S_b = {\bf \delta S}^T A {\bf \delta S} \ge 0$. We introduced the vector notation ${\bf S} := (S_0, \dots S_{k-1})^T$, and used the fact that, by definition of a PSD matrix, ${\bf u}^T A {\bf u} \ge 0$ for any vector ${\bf u}$. Therefore, for any $A \succeq 0$ and any vector ${\bf h} = (h_0, \dots h_{k-1})^T$, we have:
\begin{eqnarray}
	&{\rm Tr}(A\tilde{C}) + {\bf h} \cdot {\bf M}  = \sum_{a,b} A_{ab} \tilde{C}_{ab} + \sum_a h_a M_a \nonumber \\
	 &\ge  -\sum_{i=1}^N \left[
		\sum_{a, b} A_{ab} \langle s_a^{(i)} s_b^{(i)} \rangle \nonumber 
		 - \sum_a h_a \langle s_a^{(i)} \rangle
		 \right]\\
	 &\ge  -N E_{\rm max}(A, {\bf h}) ~, \label{eq_BI_Ct}
\end{eqnarray}
where $E_{\rm max}(A, {\bf h}) = \max_{{\bf s} \in \{\pm 1/2\}^k} [{\bf s}^T A {\bf s} - {\bf h} \cdot {\bf s}]$. This is a Bell's inequality, obeyed by all data $(C_{ab}, M_a)$ compatible with LV models, any PSD matrix $A$, and any vector ${\bf h}$. The bound $E_{\max}(A, {\bf h})$ may easily be evaluated by enumerating all $2^k$ configurations of the ${\bf s}$ variables, whenever $k$ (the number of settings) is not too large. The goal is then to find a PSD matrix $A$ and a vector ${\bf h}$ such that Eq.~\eqref{eq_BI_Ct} is violated. In order to build them, our second key observation is that $E_{\rm max}(A, {\bf h})$ is a convex function of its arguments. A simple proof of convexity, inspired by statistical physics, is to write $E_{\max}(A, {\bf h}) = \lim_{\beta \to \infty} \log Z_{\beta}(A, {\bf h})$, where $Z_\beta(A, {\bf h}) = \sum_{{\bf s} \in \{\pm 1/2\}^k} \exp[\beta ({\bf s}^T A {\bf s} -  {\bf h} \cdot {\bf s})]$, and to recognize that $\log Z_\beta$ is a convex function for any $\beta$. Furthermore, ${\rm Tr}(A\tilde{C}) + {\bf h} \cdot {\bf M}$, which is a linear function of $A$ and ${\bf h}$, is also convex. Therefore, we may introduce the convex cost function: 
\begin{equation}
L(A, {\bf h})={\rm Tr}(A\tilde{C}) + {\bf h} \cdot {\bf M} + NE_{\max}(A,{\bf h}) ~,
\label{eq_def_L_cost}
\end{equation}  which by Eq.~\eqref{eq_BI_Ct} is non-negative if $(C, {\bf M})$ are compatible with a LV model. Our data-driven algorithm \cite{code_guillem} consists therefore in solving the following optimization problem:{
\begin{eqnarray}
	&{\rm minimize}~L(A, {\bf h}) \nonumber \\
	&{\rm s.t.}~ A \succeq 0~.
	\label{eq_SDP_Ct}
\end{eqnarray}
As the PSD constraint $A \succeq 0$ maintains the convex nature of the optimization problem \cite{boyd2004convex}, if there exists a Bell's inequality of the form of Eq.~\eqref{eq_BI_Ct} which is violated by the data, then we have the guarantee to find the corresponding $A \succeq 0$ and ${\bf h}$ s.t. $L(A, {\bf h}) < 0$. Notice that if $L(A, {\bf h}) = -l < 0$ in Eq.~\eqref{eq_def_L_cost} then for any $x>0$, $L(xA, x{\bf h})=-xl$, so that $L$ is unbounded below. In a practical implementation of the algorithm, one may therefore add a cutoff on $A$ and ${\bf h}$; for this work, we have imposed $||A||^2=\sum_{a,b}A_{ab}^2 \le 1$ and $||{\bf h}||^2 = \sum_a h_a^2 \le 1$, which maintains the convex nature of the optimization. Clearly, if $L(A, {\bf h})=-l$, then by defining $x=1/\max(||A||,||{\bf h}||)$, one has  $L(xA, x{\bf h})=-xl<0$ with $||Ax||\le 1$ and $||x{\bf h}||\le 1$, and therefore adding this cutoff does not compromise the search for a violated Bell's inequality.}

Clearly, the possibility to discover new and useful Bell's inequalities via our method crucially depends on the input quantum data $\{C_{ab}, M_a\}$, which must be able to display Bell's non-locality. We will consider a situation where the quantum data are obtained by spin measurements (Sec.~\ref{sec_spin_measurements}). In this case, $C_{ab}$ and $M_a$ are completely determined by the first- and second-moments of the collective spin.
As a first application, we will recover and improve over the existing Bell's inequalities in scenarios with $d=2$ outcomes, which are violated by appropriate measurements on spin singlets \cite{frerotR2020} (Sec.~\ref{sec_BI_singlets_d2}) and spin-squeezed states \cite{schmiedetal2016} (Sec.~\ref{sec_PIBI_squeezed}), and whose violation is robust in the thermodynamic limit. In Section \ref{sec_many_outcomes}, we will then generalize these results to scenarios with $d > 2$ outcomes. 

\subsection{Spin measurements}
\label{sec_spin_measurements}
Throughout the paper, we investigate the violation of Bell's inequalities when the local measurement settings correspond to spin measurements in the $xy$-plane, in a direction independent of the subsystem. { We emphasize that this choice is only a convenient way to produce hypothetical quantum data, used as input to our data-driven algorithm. The discovered Bell's inequalities themselves are valid independently of any assumption about the system. Furthermore, we present in details several Bell's inequalities discovered by our algorithm; this lead us to derive simple conditions on the quantum state of a spin ensemble which are sufficient to violate the Bell's inequalities if the appropriate measurements are performed (in the literature, such conditions are often referred to as ``Bell-correlation witnesses''). These witness conditions are independent of the specific data we used to discover the Bell's inequalities of interest.} 
 We choose therefore:
\begin{equation}
	\hat s^{(i)}_a = \hat S_x^{(i)} \cos \theta_a + \hat S_y^{(i)} \sin \theta_a ~,
\end{equation} 
where $\hat S_x^{(i)}$ and $\hat S_y^{(i)}$ are local spin observables in directions $x$ and $y$. $\hat s^{(i)}_a$ defines a projective spin measurement along the direction $(\cos\theta_a, \sin\theta_a)$, and has therefore eigenvalues $\pm 1/2$. Introducing the collective spin $\hat J_x = \sum_{i=1}^N \hat S_x^{(i)}$ and $\hat J_y = \sum_{i=1}^N \hat S_y^{(i)}$, we also define the collective spin observables:
\begin{equation}
	\hat J_a = \sum_{i=1}^N \hat s^{(i)}_a
	 = \hat J_x \cos \theta_a + \hat J_y \sin \theta_a ~.
\end{equation}
With these conventions, the quantum data used as input of our algorithm, and against which the Bell's inequalities are evaluated, are {(for a given quantum state $\hat \rho$)}:
\begin{subequations}
\label{eq_Qdata_spin_half}
\begin{align}
	M_a = \langle \hat J_a \rangle 
	= \langle \hat J_x\rangle \cos \theta_a + \langle  \hat J_y\rangle \sin \theta_a  \\
	\tilde{C}_{ab} = \Re\langle \delta \hat J_a \delta \hat J_b \rangle - \frac{N}{4} \cos(\theta_a - \theta_b)   ~,
\end{align}
\end{subequations}
where $\delta \hat J_a = \hat J_a - M_a$, so that $\Re\langle \delta \hat J_a \delta \hat J_b \rangle = \langle \hat J_a \hat J_b + \hat J_b \hat J_a \rangle / 2 - M_a M_b$ is the covariance of the collective spin observables $\hat J_a$ and $\hat J_b$. {We defined $\langle \hat A \rangle = {\rm Tr}(\hat A \hat \rho)$, and we introduce the variance ${\rm Var}(\hat A) = \langle \hat A^2 \rangle - \langle \hat A \rangle^2$.}

\subsection{A family of Bell's inequalities for singlet-like correlations}
\label{sec_BI_singlets_d2}
As a first application of our data-driven method, we { derive Bell's inequalities maximally violated by many-body singlets, defined by $\langle \hat J_x^2\rangle = \langle \hat J_y^2\rangle = \langle \hat J_x^2\rangle = 0$. Many-body singlets are zero-eigenstates of the total spin operator $\hat{\bf J}^2=\hat J_x^2 + \hat J_y^2 + \hat J_z^2$. They are therefore $SU(2)$ invariant (that is, they are left invariant by any rotation $\exp[-i{\bf n}\cdot \hat{\bf J}]$ with ${\bf n}$ a unit vector) and generalize the Bell pair $(|\uparrow\downarrow\rangle -|\downarrow\uparrow\rangle)/\sqrt{2}$ to an arbitrary even $N$. They form a manifold of $N!/[(N/2)!(N/2+1)!]\sim N^{-3/2}2^N \sqrt{8/\pi}$ orthogonal states \cite{arecchietal1972}-- all entangled --, and are naturally produced as ground states of Heisenberg antiferromagnets \cite{Auerbach}, e.g.~at low energy in Fermi-Hubbard models \cite{TARRUELL2018365}. We emphasize that the working assumption of having a many-body singlet is only used to produce ideal quantum data, which then serve as input to our algorithm, leading us to discover new Bell's inequalities. The Bell's inequalities themselves, and the corresponding Bell correlation witnesses, are independent of any assumption about the quantum state.} It is already known that a state is entangled when ${\rm Var}(\hat J_x) + {\rm Var}(\hat J_y) < N/4$ \cite{tothetal2009}. It is also known that $[\langle \hat J_x^2 \rangle + \langle \hat J_y^2 \rangle]/N < 1 / (8 + 6 \sqrt{2}) \approx 0.060660\ldots$, which is a more demanding condition, leads to violation of a many-body Bell's inequality \cite{frerotR2020}. The measurement strategy to maximally violate the Bell's inequality of ref.~\cite{frerotR2020} is composed of $k$ coplanar spin measurements at angles $\theta_a = a\pi /k$. Our main result in this Section is to show that the Bell's inequality of ref.~\cite{frerotR2020} is not the tightest one in this measurement scenario for $k\ge 4$, leading us to discover a new family of Bell's inequalities. We find that Bell-nonlocality can be demonstrated whenever $[{\rm Var}(\hat J_x) + {\rm Var}(\hat J_y)]/N < 1/2 - 4/\pi^2 \approx 0.094715$, in the limit of $k \to \infty$.\\

\textbf{Bell's inequality.} As input quantum data, we consider a perfect spin singlet, for which $M_a = 0$ and $4\tilde{C}_{ab}/N = - \cos[\theta_a - \theta_b] = -\cos[\pi(a-b)/k]$ [from Eq.~\eqref{eq_Qdata_spin_half}{, and using the property $\hat J_a \hat\rho_{\rm singlet}=0$ for any collective spin operator $\hat J_a$ and any singlet state $\hat \rho_{\rm singlet}$}]. Applying our algorithm to these data for up to $k=10$, we find that the following Bell's inequality is violated:
\begin{subequations}\label{eq_BI_singlets_spin_half}
\begin{align}
	 \langle {\cal B} \rangle =  \sum_{a,b=0}^{k-1} \tilde{C}_{ab} \cos[\pi(a-b)/k]  \\
  \ge -N\max_{{\bf s} \in \{\pm 1/2\}^k} \sum_{a,b=0}^{k-1} s_a s_b \cos[\pi(a-b)/k] \\
  = -\frac{N}{4\sin^2[\pi/(2k)]} := B_{\rm c} ~,
\end{align}
\end{subequations}
where on the second line we used Eq.~\eqref{eq_BI_Ct}. The classical bound $B_{\rm c}$ is obtained by noting that $\sum_{a,b=0}^{k-1} s_a s_b \cos[\pi(a-b)/k] = \left\vert \sum_{a=0}^{k-1} s_a e^{ia\pi/k} \right\vert^2$. The maximum is obtained by choosing all $s_a = 1/2$, and is $1/\{4\sin^2[\pi/(2k)]\}$. \\

\textbf{Quantum violation.} To evaluate Eq.~\eqref{eq_BI_singlets_spin_half} against a generic quantum state [Eq.~\eqref{eq_Qdata_spin_half}], {not necessarily $SU(2)$ invariant,} we first introduce the matrix $A_{ab} = \cos[\pi(a-b)/k]$. Using $A_{ab} = \Re[e^{ia\pi/k} e^{-ib\pi/k}]$, the matrix $A$ is diagonalized as $A=(k/2)({\bf c} \cdot {\bf c}^T + {\bf s} \cdot {\bf s}^T)$ with the normalized vectors ${\bf c}^T = \sqrt{2/k} [\cos(a\pi/k)]_{a=0}^{k-1}$ and ${\bf s}^T = \sqrt{2/k} [\sin(b\pi/k)]_{a=0}^{k-1}$. To evaluate $\sum_{ab} \hat J_a A_{ab} \hat J_b$, we first compute $\sum_{a=0}^{k-1} c_a \hat J_a = \sqrt{2/k} \sum_{a=0}^{k-1} \cos(a\pi/k) [\cos(a\pi/k) \hat J_x + \sin(a\pi/k) \hat J_y] = \sqrt{k/2} \hat J_x$. Similarly, we find $\sum_{a=0}^{k-1} s_a \hat J_a = \sqrt{k/2} \hat J_y$. Finally, using $\sum_{a,b=0}^{k-1} \cos^2[\pi(a-b)/k] = k^2 / 2$, we find:
\begin{equation}
	\sum_{a,b=1}^{k} \tilde{C}_{ab} \cos[\pi(a-b)/k] = \frac{k^2}{4}[{\rm Var}(\hat J_x)  + {\rm Var}(\hat J_y)] -  \frac{Nk^2}{8} ~.
\end{equation}
Violation of the Bell's inequality is therefore detected whenever:
\begin{equation}
	\frac{1}{N}[{\rm Var}(\hat J_x)  + {\rm Var}(\hat J_y)] < \frac{1}{2} - \frac{1}{k^2 \sin^2(\pi / 2k)} \label{eq_condition_singlet_spin_half}~.
\end{equation}
The tightest condition is achieved in the limit $k \to \infty$, yielding the bound $1/2 - 4/\pi^2$. {This condition requires no assumption about the underlying quantum state -- apart from being composed of $N$ individual spin-$1/2$--, and only assumes a correct calibration of the measurements of $\hat J_x$ and $\hat J_y$.} Notice that the condition of Eq.~\eqref{eq_condition_singlet_spin_half} is tighter than the one derived in ref.~\cite{frerotR2020}, not only because the r.h.s is larger -- and therefore detecting more data as exhibiting Bell's non-locality in the same measurement scenario --, but also because it involves variances of the collective operators, {making the condition more robust against experimental noise -- see the related Fig.~\ref{fig_tilde_vs_nontilde}, and Section \ref{sec_experimental_noise}.}

\subsection{Spin-squeezed states}
\label{sec_PIBI_squeezed}
Spin-squeezed states of $N$ two-level systems represent paradigmatic many-body entangled states, and are a central resource for quantum-enhanced interferometry \cite{pezzeetal2016}. \\

\begin{figure}
	\includegraphics[width=1.\linewidth]{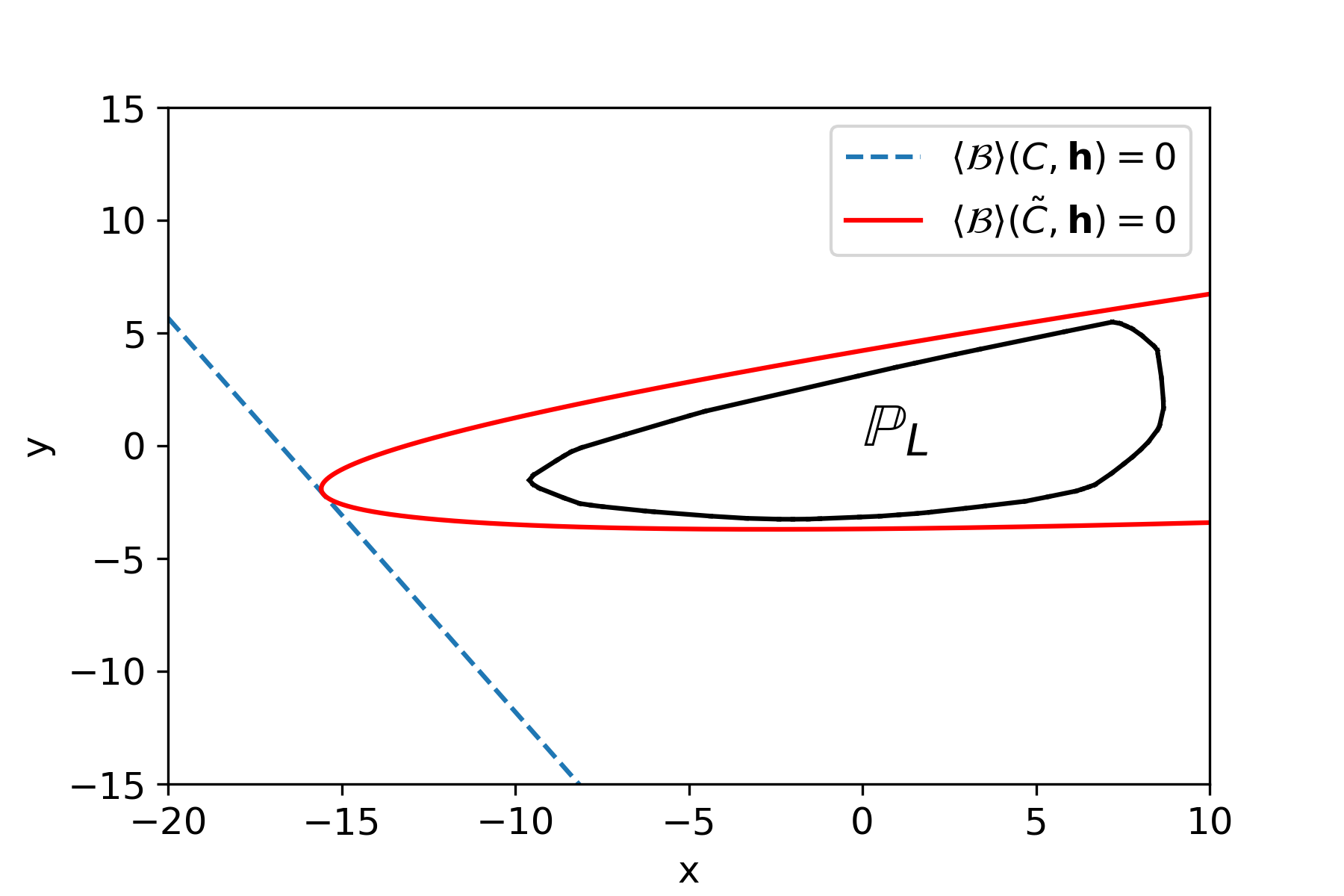}
	\caption{The (non-linear) Bell's inequalities obtained with our method, which involve $\tilde{C}_{ab}$ [Eq.~\eqref{eq_def_Ct_1}], are tighter than standard (linear) Bell's inequalities which involve $C_{ab}$ [Eq.~\eqref{eq_def_C_1}]. Here, this is illustrated for $N=10$ for the Bell's inequality of {Eq.~\eqref{eq_BI_Tura}, which strengthens the previously known Eq.~\eqref{eq_BI_Tura_0}} \cite{turaetal2014,schmiedetal2016}. We sliced the five-dimensional space $(M_0, M_1, C_{00}, C_{01}=C_{10}, C_{11})$ along a randomly-chosen $2d$ plane {[specifically, along the $({\vec u},{\vec v})$-plane with ${\vec u}=(0.07679, 0.24372, -0.34906, -0.75359, 0.49494)$ and ${\vec v}=(0.29167, -0.90583, -0.20783, -0.21443, -0.07226)$]}; $x$ and $y$ are the coordinates within this $2d$ plane. The convex black region $\mathbb{P}_L$ is the polytope of LV models, the dashed blue line is the linear Bell's inequality {of Eq.~\eqref{eq_BI_Tura_0} \cite{turaetal2014,schmiedetal2016}, and the red solid line is the corresponding non-linear Bell's inequality of Eq.~\eqref{eq_BI_Tura}, constructed from the $\tilde{C}_{ab}$ correlations, with the same coefficients.}}
	\label{fig_tilde_vs_nontilde}
\end{figure}

\textbf{State-of-the-art Bell's inequality.} In the context of Bell's non-locality, spin-squeezing is known to be essential for the robust violation of the following Bell's inequality involving $k=2$ measurement settings per subsystem \cite{turaetal2014,schmiedetal2016,engelsen_bell_2017,piga_bell_2019}:{
\begin{equation}
		\langle {\cal B} \rangle = C_{00} + C_{11} - C_{01} - C_{10} -(M_0 + M_1) \ge -N ~. \label{eq_BI_Tura_0}
\end{equation}
Notice that we are using the convention that the outcomes are $\pm 1/2$, different from the convention $\pm 1$ used in the above-cited papers; this explains why the coefficients of Eq.~\eqref{eq_BI_Tura_0} are different. We have used the experimental ref.~\cite{schmiedetal2016} to infer data serving as input to our algorithm, leading us to recover a tighter version of the above Bell's inequality:
\begin{eqnarray}\label{eq_BI_Tura}
		&\langle {\cal B} \rangle = \tilde{C}_{00} + \tilde{C}_{11} - \tilde{C}_{01} - \tilde{C}_{10} -(M_0 + M_1)  \nonumber \\
		 &= C_{00} + C_{11} - C_{01} - C_{10} -(M_0-M_1)^2 -(M_0 + M_1)\nonumber \\
		 &\ge -N~. 
\end{eqnarray}
 This shows that, if Eq.~\eqref{eq_BI_Tura_0} had not been known from ref.~\cite{turaetal2014}, the tighter Eq.~\eqref{eq_BI_Tura} would have been recovered \textit{in a data-driven way} by our method. This clearly demonstrates the concrete advantage offered by our method in analyzing experimental data in an unbiased way.
Notice that while the coefficients of the Bell's inequalities Eq.~\eqref{eq_BI_Tura_0} and Eq.~\eqref{eq_BI_Tura} are the same, Eq.~\eqref{eq_BI_Tura_0} involves the correlations $C_{ab}$ [Eq.~\eqref{eq_def_C_1}], and not $\tilde{C}_{ab} = C_{ab} - M_a M_b$ [Eq.~\eqref{eq_def_Ct_1}]. Therefore, Eq.~\eqref{eq_BI_Tura} includes the extra term $-(M_0 - M_1)^2 \le 0$, and is therefore strictly tighter than Eq.~\eqref{eq_BI_Tura_0} -- see Fig.~\ref{fig_tilde_vs_nontilde} for an illustration. Since this extra term is of order $O(N^2)$ while the classical bound is $O(N)$, the relative improvement generically grows with $N$. 
}
The classical bound is found, following Eq.~\eqref{eq_BI_Ct}, by writing ${\cal B} = (\delta S_0 - \delta S_1)^2 - \sum_{i=1}^N[(s_0 - s_1)^2 + s_0 + s_1]^{(i)}$, and noting that $(s_0 - s_1)^2 + s_0 + s_1 \ge -1$ for all possibles values of $s_0, s_1 = \pm 1/2$. This Bell's inequality can be violated by preparing a spin-squeezed state, defined by $N{\rm Var}(\hat J_y) < \langle \hat J_x \rangle^2$ \cite{pezzeetal2016}, and performing two projective spin measurements in directions $\hat s^{(i)}_a = \hat S_x^{(i)} \cos \theta \pm \hat S_y^{(i)} \sin \theta$ \cite{schmiedetal2016,piga_bell_2019}. Computing the quantum value from Eq.~\eqref{eq_Qdata_spin_half}, we obtain $\langle {\cal B} \rangle = 4 \sin^2 \theta {\rm Var}(\hat J_y) - 2\cos \theta \langle J_x \rangle - N \sin^2\theta $. The optimal angle $\theta$, minimizing $\langle {\cal B}\rangle$ for fixed data (${\rm Var}(\hat J_y), \langle \hat J_x \rangle$), is $\cos \theta = \langle \hat J_x \rangle / [N - 4{\rm Var}(\hat J_y)]$. For this choice of measurements, we obtain $\langle B \rangle = -N + 4{\rm Var}(\hat J_y) - \langle \hat J_x \rangle^2 / [N - 4 {\rm Var}(\hat J_y)]$. { Notice that for Eq.~\eqref{eq_BI_Tura_0}, a similar condition may be derived, but involving $\langle \hat J_y^2\rangle$ instead of ${\rm Var}(\hat J_y)$. Whenever $\langle \hat J_y \rangle=\epsilon N$ with $\epsilon \neq 0$, $\langle \hat J_y^2 \rangle \sim N^2$, which represents a fundamental obstruction to the violation of Eq.~\eqref{eq_BI_Tura_0} in the thermodynamic limit for non-ideal data. Instead, working with the tighter Eq.~\eqref{eq_BI_Tura}, and the corresponding criterion involving ${\rm Var}(\hat J_y)$, such obstruction is removed.} For perfect squeezed states [${\rm Var}(\hat J_y) \to 0$, $\langle \hat J_x \rangle \to N/2$], we can obtain violation up to $\langle {\cal B} \rangle = -5N/4$.
 In this Section, we show that the robustness of Bell non-locality detection for spin-squeezed states can be improved by considering extra measurements ($k \ge 3$). \\
 
\begin{figure}
	\includegraphics[width=\linewidth]{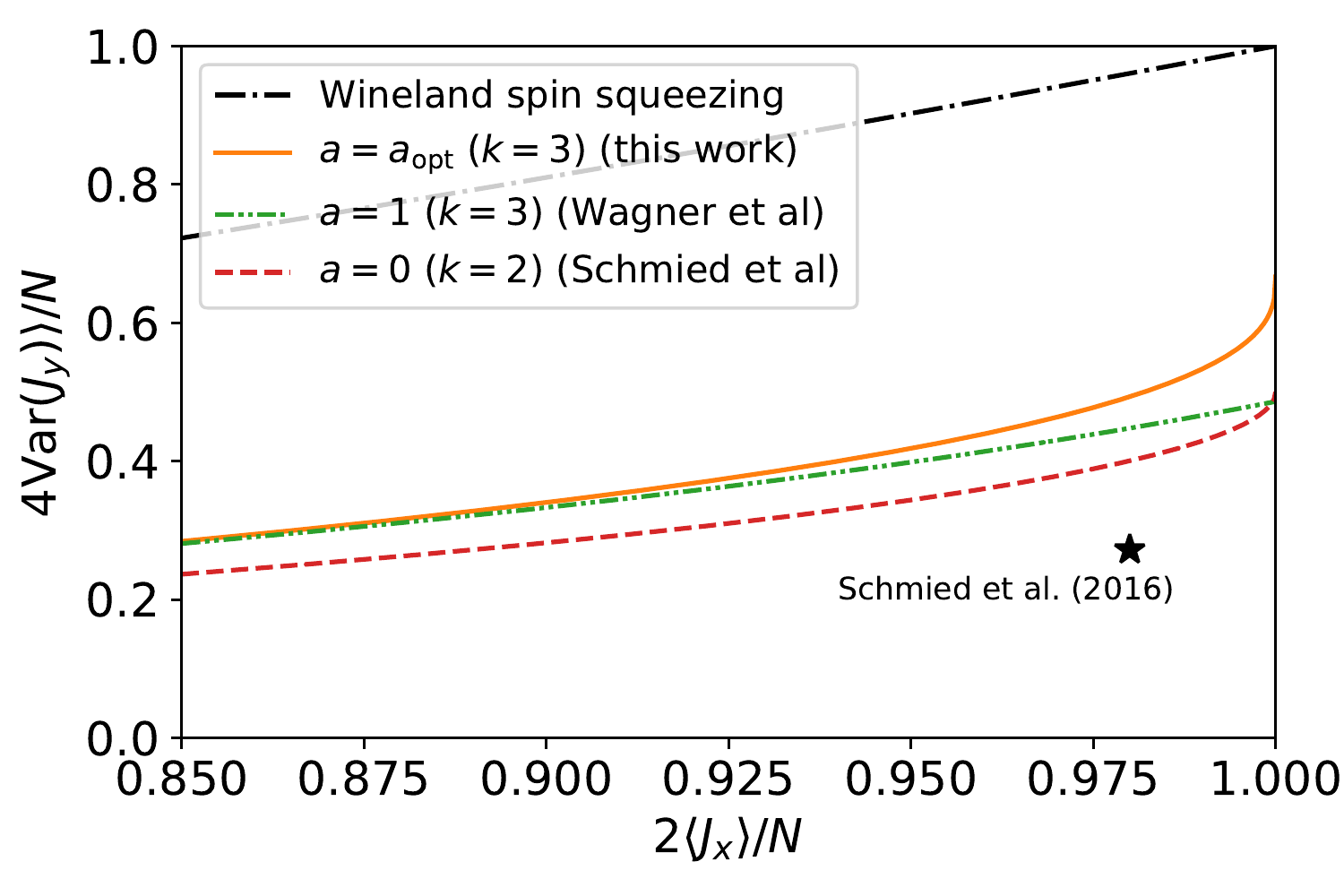}
	\caption{Entanglement detection for $j=1/2$ spin-squeezed states. Below each line, the corresponding entanglement criterion is violated. { Dashed-dotted black line: Wineland spin squeezing criterion; solid orange line: violation of the Bell's inequality Eq.~\eqref{eq_BI_Turalike_k3} with the optimal parameter $a$; dashed-dotted-dotted green line: same Bell's inequality with $a=1$ \cite{wagner_bell_2017}; dashed red line: violation of the Bell's inequality Eq.~\eqref{eq_BI_Tura} \cite{schmiedetal2016}. Black star: experimental data from ref.~\cite{schmiedetal2016}, assuming $\langle \hat J_y \rangle=0$. Notice that the previously known results were in fact involving $\langle \hat J_y^2 \rangle$. We have proved in this work that they remain valid with ${\rm Var}(\hat J_y)$ instead of $\langle  \hat J_y^2 \rangle$, leading to systematically tighter criteria. In particular, the data of ref.~\cite{schmiedetal2016} (black star), using the actual value of $\langle  \hat J_y \rangle$ in the experiment, would be lower along the vertical axis, and similarly for the data of ref.~\cite{engelsen_bell_2017} (not shown)}.} 
	\label{fig_new_PIBI_squeezed_state_phase_diagram}
\end{figure}

 \textbf{Finding tightest and more robust Bell's inequalities.}
 \begin{figure}
	\includegraphics[width=\linewidth]{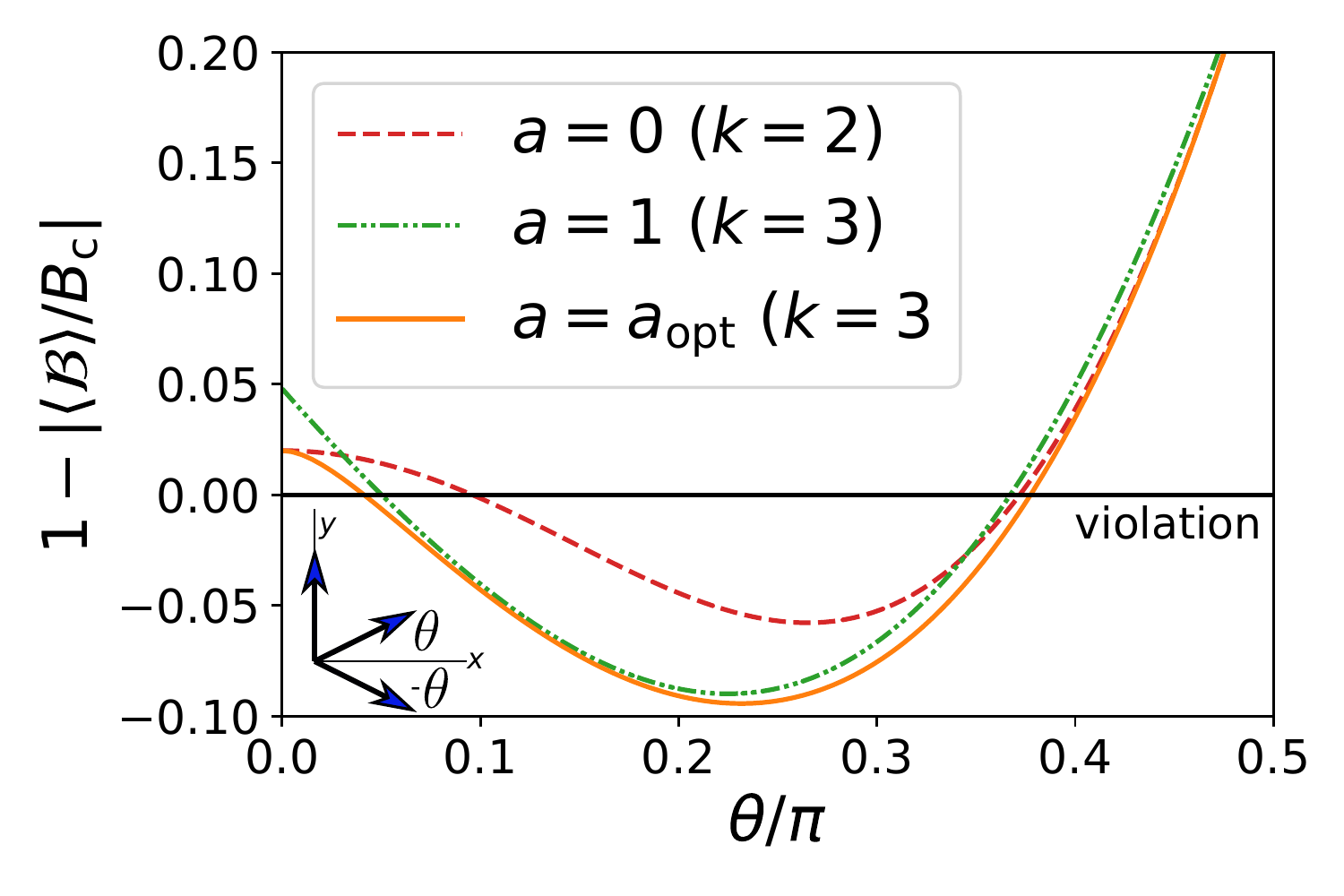}
	\caption{Bell's inequality violation for $j=1/2$ spin-squeezed states, as a function of the measurement angle $\theta$. Dashed red line: relative violation of the Bell's inequality of Eq.~\eqref{eq_BI_Tura}, which involves $k=2$ measurement settings at angles $\pm \theta$. { Dashed-dotted-dotted green line: relative violation of the Bell's inequality of Eq.~\eqref{eq_BI_Turalike_k3}, which involves a third measurement along the $y$-axis (see the sketch on the bottom-left corner), with $a=1$ in Eq.~\eqref{eq_BI_Turalike_k3}. Solid orange line: same inequality, for the optimal value of the parameter $a$. The quantum state, chosen from ref.~\cite{schmiedetal2016}, has a mean spin $2\langle \hat J_x \rangle / N = m_x = 0.98$, and transverse collective-spin fluctuations $4{\rm Var}(\hat J_y) / N = \chi^2 = 0.272$ (assuming $\langle \hat J_y \rangle=0$)}. For all inequalities, the absolute value of the relative violation is equal to the amount of white noise tolerated by the data to observe a non-zero violation (see text).}
	\label{fig_new_PIBI_squeezed_state}
\end{figure}
 To find better Bell's inequalities, our strategy was to consider quantum data [Eq.~\eqref{eq_Qdata_spin_half}] obtained from a squeezed state at the limit of violating the Bell's inequality Eq.~\eqref{eq_BI_Tura}, and add extra measurements in the $xy$-plane to potentially discover other violated Bell's inequalities. In particular, adding a third spin measurement $\hat S_y^{(i)}$ along the $y$-axis, we found a family of Bell's inequalities, defined by the following coefficients [see Eq.~\eqref{eq_BI_Ct}]:
 \begin{subequations}
 \begin{align}
	&A = \begin{pmatrix}
	1 & -1 & a \\
	-1 & 1 & -a \\
	a & -a & a^2\\	
\end{pmatrix} \\
	&{\bf h}^T = -(1+a, 1+a, 0) ~,
  \end{align}
\end{subequations}
where $a \ge 0$.
The corresponding Bell's inequality reads:
\begin{equation}
	\langle {\cal B} \rangle = {\rm Tr}(A \tilde{C}) + {\bf h} \cdot {\bf M} \ge -N(1 + a/2)^2 := B_{\rm c}~,
	\label{eq_BI_Turalike_k3}
\end{equation}
and reduces to Eq.~\eqref{eq_BI_Tura} when $a=0$.
{Remarkably, for $a=1$, Eq.~\eqref{eq_BI_Turalike_k3} represents a tighter version of a Bell's inequality analyzed in ref.~\cite{wagner_bell_2017} -- tighter, due to the non-linear nature of Eq.~\eqref{eq_BI_Turalike_k3} which involves $\tilde{C}$ instead of $C$. Similarly to Eq.~\eqref{eq_BI_Tura}, we discovered this family of Bell's inequalities parametrized by $a$ in a data-agnostic way, using data from a spin-squeezed state as input to our algorithm.
The classical bound $B_{\rm c}$ may be found in the following manner.} Noting that ${\bf x}^T A {\bf x} = (x_0 - x_1 + ax_2)^2$ for any vector ${\bf x}$, we may write:
\begin{eqnarray}
	{\cal B} =  [\delta(S_0 - S_1 + a S_2)]^2 - 
	\sum_{i=1}^N \{ \nonumber \\ 
		(1+a)[s_0 + s_1] +
		[s_0 - s_1 + a s_2]^2
		\}^{(i)} ~.
\end{eqnarray}
The classical bound $B_{\rm c} = -N(1 + a/2)^2$ is then found by enumerating the configurations of the variables $s_a^{(i)}=\pm 1/2$. On the other hand, in the quantum measurement setting we consider, we find:
\begin{equation}
	\langle {\cal B} \rangle = [{\rm Var}(\hat J_y) - N/4](a + 2 \sin \theta)^2 - 2(1+a) \langle \hat J_x \rangle \cos \theta ~.
\end{equation}
The quantum data $(\tilde{C}_{ab}, M_a)$ being fixed, it is then natural to consider the optimal values of $\theta$ and $a$ to have the most robust violation of the Bell's inequality. If white noise is added to the data, then $({C}_{ab}, M_a) \to (1-r)({C}_{ab}, M_a)$ with $r$ the noise amplitude. If we assume that $\langle \hat J_y \rangle = 0$, so that ${\rm Var}(\hat J_y) = \langle \hat J_y^2 \rangle$, then correspondingly $\langle {\cal B} \rangle \to (1-r) \langle {\cal B} \rangle$. {The noise robustness may be interpreted as the intrinsic robustness of a given Bell's inequality violation against generic errors during the preparation of the quantum system, modelled as $\hat \rho = (1-r)\hat \rho^{(\rm ideal)} + r \mathbb{1}/D$ with $D$ the dimension of the total Hilbert space (a more detailed discussion of the experimental requirements to accurately estimate the data is given in Section \ref{sec_experimental_noise}).} Maximizing the noise robustness is therefore equivalent to maximizing the ratio $|\langle {\cal B} \rangle / B_{\rm c}| = [(1+a) m_x \cos \theta + (1 - \chi^2)(a / 2 + \sin \theta)^2] / (1 + a/2)^2$, where we introduced $m_x = 2\langle \hat J_x \rangle / N$ (the Rabi contrast \cite{schmiedetal2016}) and $\chi^2 = 4{\rm Var}(\hat J_y)/N$ the scaled second moment. For each value of $\theta$, we may then find the value of $a$ which maximizes this ratio. As illustrated on Fig.~\ref{fig_new_PIBI_squeezed_state} for the data of ref.~\cite{schmiedetal2016} ($m_x=0.98$ and $\chi^2=0.272$ { assuming that $\langle \hat J_y \rangle=0$), adding a third measurement along the $y$-axis and optimizing over the parameter $a$ yields a systematic improvement over both Eq.~\eqref{eq_BI_Tura} (which involves only two measurement settings), and over Eq.~\eqref{eq_BI_Turalike_k3} with $a=1$, as proposed in ref.~\cite{wagner_bell_2017}. Notice that we assumed that $\langle \hat J_y \rangle=0$. Therefore, in the specific measurement settings we considered, working with the $\tilde{C}$ quantities in the Bell's inequalities Eq.~\eqref{eq_BI_Tura} or Eq.~\eqref{eq_BI_Turalike_k3} is equivalent to $C$; however, in analyzing experimental data where $\langle \hat J_y \rangle$ is never exactly zero, the non-linear nature of our Bell's inequalities (namely, working with $\tilde{C}$) will lead to a systematic improvement over all previously known Bell's inequalities \cite{turaetal2014,schmiedetal2016,engelsen_bell_2017,wagner_bell_2017}. It would be interesting to analyze the experimental data of refs.~\cite{schmiedetal2016,engelsen_bell_2017} from this perspective -- this would certainly lead to a more robust detection of Bell correlations.

Finally, for given values of $(m_x, \chi^2)$, we may find the measurement angle $\theta$ and parameter $a$ which maximize the robustness of the violation of Eq.~\eqref{eq_BI_Turalike_k3}. This is done on Fig.~\ref{fig_new_PIBI_squeezed_state_phase_diagram}, which shows the parameter regime in the $(m_x, \chi^2)$ where Bell non-locality is detected. For comparison we also plot the regime where non-locality is detected based on the violation of Eq.~\eqref{eq_BI_Tura}, on the violation of Eq.~\eqref{eq_BI_Turalike_k3} with $a=1$, and where entanglement is detected based on the Wineland spin squeezing criterion $m_x^2 > \chi^2$. The Bell's inequality Eq.~\eqref{eq_BI_Turalike_k3} with optimal $a$ systematically extends the parameter space where non-locality can be detected with $k=3$ settings. Notice that this parameter space can be further extended by considering more measurement settings \cite{wagner_bell_2017}.}\\

\textbf{Further improvement.}
Even more robust Bell's inequalities may be found by considering $k\ge 2$ pairs of measurements in the $xy$ plane, in directions $\hat s_{a}^{(i)} = \hat S_x^{(i)} \cos \theta_a + \hat S_y^{(i)} \sin \theta_a$ and $[\hat s_{a}^{(i)}]' = \hat S_x^{(i)} \cos \theta_a - \hat S_y^{(i)} \sin \theta_a$, and one measurement along $y$ denoted $\hat s_k^{(i)} = \hat S_y^{(i)}$. In this Bell scenario with $2k+1$ settings applied to spin-squeezed states, this generically leads to Bell' inequalities of the form:
\begin{eqnarray}
	\left[\sum_{a=0}^{k-1} \alpha_a(S_a - S_a') + S_k \right]^2 - \sum_{i=1}^N \{ \nonumber \\
	 \sum_{a=0}^{k-1} \beta_a(s_a + s_a') +
	  \left[\sum_{a=0}^{k-1} \alpha_a(s_a - s_a') + s_k \right]^2 
	   \}^{(i)} \nonumber \\ 
	   \ge B_{\rm c}
\end{eqnarray}
for some data-tailored coefficients $\alpha_a$ and $\beta_a$. Similarly as in the previous paragraph, for given values of $(m_x, \chi^2)$, it is then possible to numerically optimize over the measurement angles $\theta_a$ in order to maximize the violation robustness. { Similarly to the case with three measurements discussed above, one can expect to obtain systematically tighter Bell's inequalities as compared to ref.~\cite{wagner_bell_2017}, where an arbitrary number of settings are considered in a similar measurement scenario.}

\section{Arbitrary-outcomes measurements}
\label{sec_many_outcomes}
\textbf{Summary of the main results.}
In the previous section, we focused on measurements with $d=2$ outcomes -- and considered a physical implementation with spin-$1/2$ measurements. In this section, we extend these results to arbitrarily-many outcomes ($d>2$), corresponding to the physical situation where spin measurements are performed on individual spin-$j$ components (with $d=2j+1$). Sec.~\ref{sec_algo_spin_j} presents an incremental generalization of the algorithm of Sec.~\ref{sec_convex_opt_algo_d2}, which incorporates an extra feature of the quantum data [Eq.~\eqref{Qdata_Ma2}]. This turns out to be an essential ingredient to generalize the Bell's inequalities of the previous section. We first consider spin-$j$ singlets in Sec.~\ref{sec_singlets_spin_j}, and restrict our attention to $k=3$ spin measurements in a given plane. We unveil an increasingly complex situation for $j>1/2$, as illustrated on Fig.~\ref{fig_zoo}, with $2j$ inequivalent Bell's inequalities, already in this simple setting. We could however characterize analytically two families of Bell's inequalities which emerged from our algorithm (Sec.~\ref{sec_singlets_spin_j}). One of them extends Eq.~\eqref{eq_BI_singlets_spin_half} to arbitrary half-integer spins (for $k=3$ measurements), and the corresponding witness condition is given by Eq.~\eqref{eq_criterion_singlets_1}. The other family is valid for both integer and half-integer spins, and the witness condition is given by Eq.~\eqref{eq_criterion_singlets_2}. We conclude our exploration in Sec.~\ref{sec_PIBI_squeezed} with spin-$j$ squeezed states. Our main result is a generalization of the Bell's inequality of Eq.~\eqref{eq_BI_Tura} to arbitrary $j\ge 1/2$ [Eq.~\eqref{eq_BI_Tura_spin_j}]. The corresponding witness condition for spin-$j$ squeezed states is given by Eq.~\eqref{eq_witness_squeezing_spin_j}. 

\subsection{Algorithm tailored to spin measurements}
\label{sec_algo_spin_j}
We consider the general scenario in which the local measurements $\hat s_a^{(i)}$ ($a \in  \{0, \dots k-1\}$, $i \in \{1, \dots N\}$) can deliver $d \ge 2$ possible outcomes, denoted $s = \{-j, -j+1, \dots j\}$ with $d=2j+1$. In general, the pair probability distribution $P^{(ij)}(s,t|a,b)$ may be reconstructed from the single-body expectation values $\langle [s_a^{(i)}]^\alpha \rangle$ with $\alpha \in \{1, \dots d-1\}$, and two-body correlations $\langle [s_a^{(i)}]^\alpha [s_b^{(j)}]^\beta \rangle$ with $\alpha, \beta \in \{1, \dots d-1\}$. The averaged pair probability distribution $\bar{P}(s,t|a,b) = [N(N-1)]^{-1} \sum_{i \neq j} P^{(ij)}(s,t|a,b)$ may then be obtained by averaging over all permutations of the subsystems. In Appendix \ref{sec_app_algo}, we give a general formulation of our data-driven algorithm for finding a Bell's inequality violated by $\bar P$. However, aiming at finding new Bell's inequalities for many-spin systems with $j > 1/2$, we found sufficient to include only:
\begin{equation}
	M_a^{(2)} = \sum_{i=1}^N \langle [s_a^{(i)}]^2 \rangle
	\label{Qdata_Ma2}
\end{equation}
as an extra coarse-grain feature of the quantum data, in addition to $M_a$ and $C_{ab}$ defined in Eq.~\eqref{eq_def_C}. Apart from this modification, we may then follow the same construction as in Section \ref{sec_convex_opt_algo_d2}, where the $kN$ local classical variables $s_a^{(i)}$ can now take the $d$ possible values $-j, -j+1, \dots j$. The analogue of Eq.~\eqref{eq_BI_Ct} now contains an extra term ${\bf h}^{(2)} \cdot {\bf M}^{(2)}$ to allow for Bell's inequalities involving this extra feature of the data. Explicitly, for any PSD matrix $A$, and any vectors ${\bf h}=(h_1, \dots h_k)$ and ${\bf h}^{(2)} = (h_1^{(2)}, \dots h_k^{(2)})$, we have:
\begin{eqnarray}
	&{\rm Tr}(A\tilde{C}) + {\bf h} \cdot {\bf M} + {\bf h}^{(2)} \cdot {\bf M}^{(2)} \nonumber \\
	& = \sum_{a,b} A_{ab} \tilde{C}_{ab} + \sum_a h_a M_a + \sum_a h_a^{(2)} M_a^{(2)} \nonumber \\
	 &\ge  -\sum_{i=1}^N \left\langle
		\sum_{a, b} A_{ab}  s_a^{(i)} s_b^{(i)} 
		 - \sum_a h_a s_a^{(i)} - \sum_a h_a^{(2)} [s_a^{(i)}]^2 
		 \right\rangle\nonumber \\
	 &\ge  -N E_{\rm max}(A, {\bf h}, {\bf h}^{(2)}) ~, \label{eq_BI_Ct_arbitrary_d}
\end{eqnarray}
where now $E_{\rm max}(A, {\bf h}, {\bf h}^{(2)}) = \max_{{\bf s} \in \{-j, \dots j\}^k} E({\bf s})$, with $E({\bf s}) =  \sum_{ab} A_{ab} s_a s_b - \sum_a [h_a s_a + h_a^{(2)} s_a^2]$. Eq.~\eqref{eq_BI_Ct_arbitrary_d} is a Bell's inequality, satisfied by all data $M_a$, $M_a^{(2)}$ and $C_{ab}$ compatible with a LV model with $d$-outcome measurements. We may then parallel the end of Section \ref{sec_convex_opt_algo_d2}: introduce the convex cost function $L(A, {\bf h}, {\bf h}^{(2)})={\rm Tr}(A\tilde{C}) + {\bf h} \cdot {\bf M} + {\bf h}^{(2)} \cdot {\bf M}^{(2)} + NE_{\max}(A,{\bf h}, {\bf h}^{(2)})$, and minimize it via a convex-optimization routine, imposing the PSD constraint $A \succeq 0$. If we find $L<0$, a violated Bell's inequalities is then reconstructed from the corresponding $A$, ${\bf h}$ and ${\bf h}^{(2)}$. 

Applying this algorithm, we discovered Bell's inequalities violated by spin-$j$ spin singlets, and by spin-$j$ squeezed states. These Bell's inequalities generalize the results of Section \ref{sec_d2} to arbitrary $j \ge 1/2$. 

\begin{figure}
\includegraphics[width=1.\linewidth]{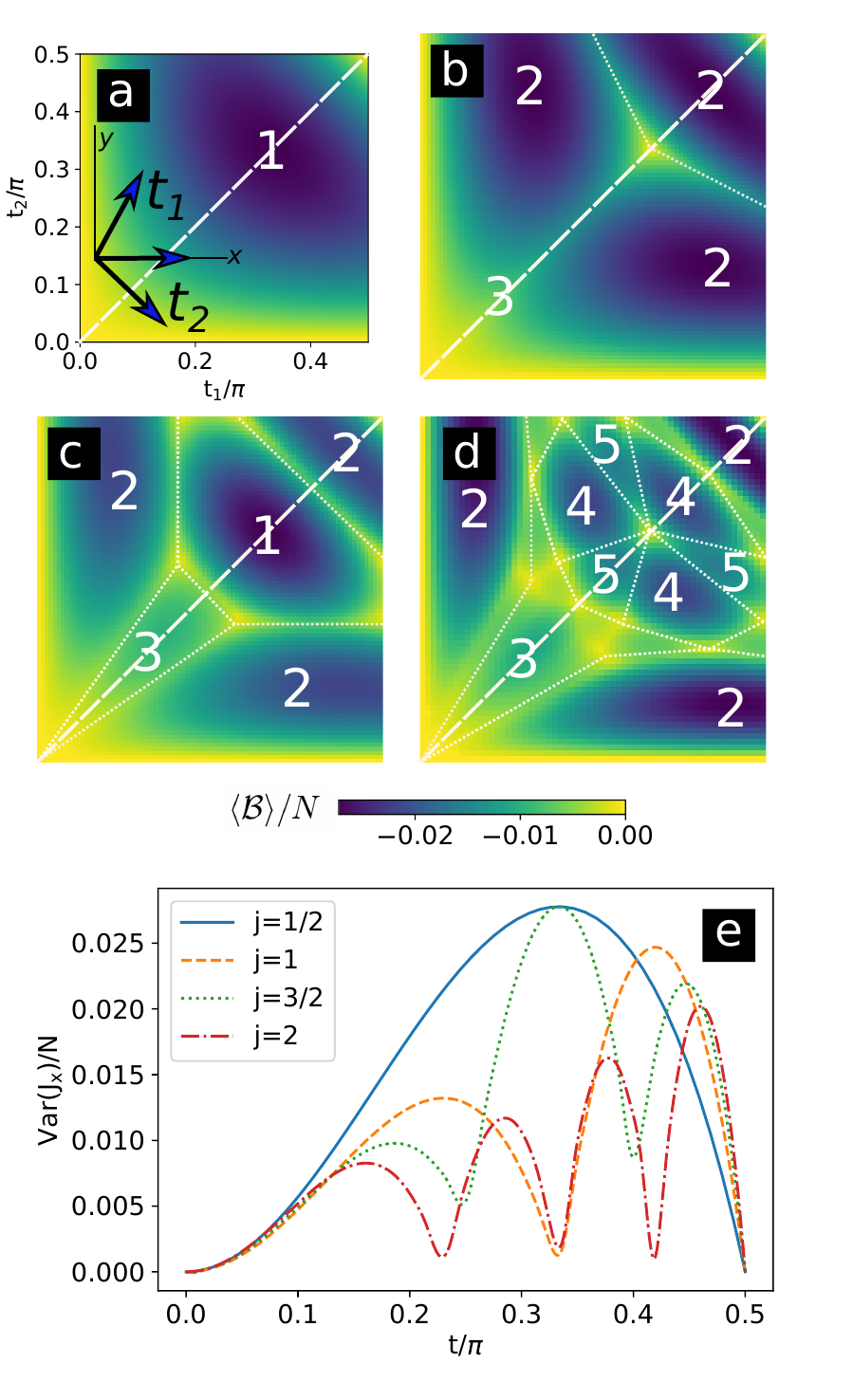}
\caption{Bell's inequalities for many-body spin singlets. As input quantum data [Eqs.~\eqref{eq_Qdata_singlets}] to our algorithm, we considered $k=3$ spin measurements in the $xy$-plane, forming angles $t_1$, $0$, and $-t_2$ with the $x$ axis {(inset of panel a)}. We found violated Bell's inequalities as in Eq.~\eqref{eq_BI_SU2_generic}, with a conventional normalization $\sum_{ab} A_{ab}^2 + \sum_a [h_a^{(2)}]^2= 1$, and a classical bound given by Eq.~\eqref{eq_Emax_singlets}. (a,b,c,d): Difference between the classical bound and the quantum value [Eq.~\eqref{eq_BQ_singlets_1}], divided by $N$. (a) $j=1/2$; (b) $j=1$; (c) $j=3/2$; (d) $j=2$. The integers $1-5$ label inequivalent families of Bell's inequalities, which are found in the different regions of parameters $(t_1,t_2)$ (boundaries between these regions are indicatively emphasized as dotted white lines). In general, we find $2j$ Bell's inequalities inequivalent under relabelling of the outcomes. Labels $1$ and $2$ respectively correspond to the coefficients of Eqs.~\eqref{eq_Atilde_h2_1} and \eqref{eq_Atilde_h2_2}. (e) Along $t:=t_1=t_2$ [white dashed line on panels (a-d)], bound on ${\rm Var}(\hat J_x)/N$ to observe Bell non-locality assuming $SU(2)$ invariance in this measurement setting [Eq.~\eqref{eq_witness_SU2}]. The maximum at $t=\pi / 3$ for half-integer spins corresponds to label $1$, and the witness condition is that of Eq.~\eqref{eq_criterion_singlets_1}. The right-most local maximum at $t = \arccos[1/(4j)]$ corresponds to label $2$, and the witness condition is that of Eq.~\eqref{eq_criterion_singlets_2}.
}
\label{fig_zoo}
\end{figure}

{
\subsection{Bell's inequalities for arbitrary-$j$ many-body singlets}
\label{sec_singlets_spin_j}
We start our investigation of Bell's inequalities tailored to spin-$j$ systems by considering, as input to our algorithm, many-body singlets. This will lead us to extend some the results obtained in Section \ref{sec_BI_singlets_d2} for $j=1/2$ to arbitrary spins. Here again, the assumption of having a many-body singlet is only used to produce quantum data leading us to discover new Bell's inequalities via our data-driven algorithm. The Bell's inequalities are independent of any assumption about the quantum systems being measured, and the witness inequalities only assume that a collection of $N$ spin-$j$ particles are measured along appropriately-calibrated axes. As in the case of $j=1/2$, spin singlets are $SU(2)$-invariant states defined by the sole condition $\langle \hat J_x^2 \rangle = \langle \hat J_y^2 \rangle = \langle \hat J_z^2 \rangle = 0$. It is known that if ${\rm Var}(\hat J_x) + {\rm Var}(\hat J_y) + {\rm Var}(\hat J_z) \le Nj$, then the state is multipartite entangled \cite{vitaglianoetal2011}, and therefore spin singlets are entangled for any $j$. We have considered $k=3$ coplanar spin measurements, in directions $\hat s_a^{(i)} = \hat S_x^{(i)} \cos \theta_a + \hat S_y^{(i)} \sin \theta_a$, with $\{\theta_a\}=(t_1, 0, -t_2)$.} We did not find violated Bell's inequalities with $k=2$ settings, and using non-coplanar spin measurements did not lead to more robust Bell's inequalities. However, we do not exclude that better Bell's inequalities could be found using non-coplanar measurements, with $k\ge 4$ settings, or including more general $SU(2j+1)$ measurements. In summary, this setting appeared as the simplest one to discover new Bell's inequalities, and even in this simplest scenario we could not characterize all the Bell's inequalities which appear when increasing $j$. Fig.~\ref{fig_zoo} summarizes our findings, where we plot the violation of Bell's inequalities in the $(t_1, t_2)$ plane, for $j\in \{1/2,1,3/2,2\}$, together with the witness condition on the collective spin variance ${\rm Var}(\hat{J_x})$ to observe violation (assuming global $SU(2)$ invariance). We shall discuss in details two families of Bell's inequalities, which were characterized analytically for arbitrary $j$.\\

\subsubsection{General considerations}
{ We begin with general considerations on the Bell's inequalities discovered by using, as input to our algorithm, the data obtained measuring a many-body singlet along $k$ directions in the $xy$-plane. As a consequence of $SU(2)$ invariance, a many-body singlets has no mean spin orientation: for any direction $a$, $\langle \hat J_a \rangle = \sum_{i=1}^N \langle \hat s_a^{(i)}\rangle = 0$. Furthermore, as a consequence of $[\hat S_x^{(i)}]^2 + [\hat S_y^{(i)}]^2 + [\hat S_z^{(i)}]^2 = j(j+1)$ and of $SU(2)$ invariance, we have $\sum_{i=1}^N \langle [\hat s_a^{(i)}]^2\rangle = Nj(j+1)/3$. In a many-body singlet, the quantum data $\tilde{C}_{ab}$, $M_a$ and $M_a^{(2)}$ are [Eqs.~\eqref{eq_def_Ct} and \eqref{Qdata_Ma2}]:
\begin{subequations}
 \label{eq_Qdata_singlets}
 \begin{align}
	&\tilde{C}_{ab} = - [Nj(j+1)/3] \cos(\theta_a - \theta_b) 		\label{eq_Qdata_singlets_1} \\
	&M_a = 0 \label{eq_Qdata_singlets_2}\\
	&M_a^{(2)} = Nj(j+1) / 3 \label{eq_Qdata_singlets_3}~,
 \end{align}
\end{subequations}
(in the specific cases discussed in further details below, $\theta_0 = t_1$, $\theta_1 = 0$ and $\theta_2 = -t_2$).} \\

\noindent\textbf{Structure of the Bell's inequalities.}
As a consequence of $M_a=0$, the Bell's inequalities tailored to singlets do not involve terms linear in $M_a$ [namely, in the notations of Eq.~\eqref{eq_BI_Ct_arbitrary_d}, we have ${\bf h} = 0$], and they take the general form:
\begin{equation}
	\langle {\cal B} \rangle = \sum_{ab} A_{ab} \tilde{C}_{ab} + \sum_a h_a^{(2)} M_a^{(2)} \ge B_{c} ~,
\label{eq_BI_SU2_generic}
\end{equation}
where $A$ is a symmetric PSD matrix. 
For LV models, we have:
\begin{eqnarray}
	{\cal B}_{\rm LV} &=& \sum_{ab} \delta S_a A_{ab} \delta S_b - \sum_i \sum_{ab} s^{(i)}_a [A_{ab} - h_a^{(2)} \delta_{ab}] s^{(i)}_b \nonumber \\
	&\ge& - NE_{\rm max}(\tilde{A}) ~,\label{eq_bell_operator_singlets_j}
\end{eqnarray}
where we defined $\tilde A_{ab} = A_{ab} - h_a^{(2)} \delta_{ab}$, and:
\begin{equation}
	E_{\rm max}(\tilde{A}) := \max_{\{s_a\} \in \{-j, \dots j\}^k} \sum_{ab} s_a \tilde{A}_{ab} s_b ~.
	\label{eq_Emax_singlets}
\end{equation}
Notice that the bound is tight for $N$ even. Indeed, if $\{s_a^{\rm (opt)}\}_{a=0}^{k-1}$ is a configuration saturating $E_{\max}$, then $\{-s_a^{\rm (opt)}\}_{a=0}^{k-1}$ is also saturating $E_{\max}$. We may therefore always choose $S_a = \sum_{i=1}^N s_a^{(i)} = 0$, while saturating the bound, by choosing the configuration $\{s_a^{\rm (opt)}\}_{a=0}^{k-1}$ for $N/2$ subsystems, and $\{-s_a^{\rm (opt)}\}_{a=0}^{k-1}$ for the other $N/2$ subsystems. \\

{\noindent\textbf{Quantum value on a singlet.}
Considering the quantum value on a spin singlet [Eq.~\eqref{eq_Qdata_singlets}], we find:
\begin{equation}
	\langle {\cal B} \rangle_{\rm singlet} = - \frac{Nj(j+1)}{3} \sum_{ab} \tilde A_{ab} \cos(\theta_a - \theta_b)
	\label{eq_BQ_singlets_1}
\end{equation}
Using that $\tilde A_{ab} = \tilde A_{ba}$, we have $\sum_{ab} \tilde A_{ab} \cos(\theta_a - \theta_b) = \sum_{ab} e^{-i\theta_a} \tilde{A}_{ab} e^{i\theta_b}$.} The optimal angles, leading to the maximal violation of the Bell's inequality, are those which maximize $\sum_{ab} e^{-i\theta_b} \tilde{A}_{ab} e^{i\theta_a}$. We therefore define:
\begin{equation}
	Q_{\rm max}(\tilde{A}) := \max_{\{\theta_a\} \in [-\pi, \pi]^k} \sum_{ab} e^{-i\theta_a} \tilde{A}_{ab} e^{i\theta_b} ~.
	\label{eq_Qmax_singlets}
\end{equation}
 This should be contrasted to the case of LV models where, in order to find the classical bound, one maximizes $\sum_{ab} s_a \tilde{A}_{ab} s_b$ over the variables $s_a \in \{-j, \dots j\}$ [see Eq.~\eqref{eq_Emax_singlets}]. On Fig.~\ref{fig_zoo}(a-d), we plot the quantum violation of the Bell's inequalities we found, for $j\in \{1/2,1,3/2,2\}$, with $k=3$ spin-measurement directions $\{\theta_a\} = (t_1, 0, -t_2)$. In general, varying the measurement angles $t_1$ and $t_2$, we find $2j$ inequivalent inequalities. We characterized analytically one Bell's inequality appearing for all half-integer $j$, and one family appearing for all $j$ (see below).
 \\
 
{\noindent\textbf{Witness condition.}
For each Bell's inequality of the form Eq.~\eqref{eq_BI_SU2_generic} found by our approach, and for given measurement directions $\{\theta_a\}$, one may derive witness conditions which can be measured via global measurements on an ensemble of $N$ spin-$j$ particles, and which demonstrate the capability of the quantum state to violate the considered Bell's inequality without further assumptions (in particular, without assuming $SU(2)$ invariance). We first express the average value of the Bell operator [Eqs.~\eqref{eq_BI_SU2_generic} and \eqref{eq_bell_operator_singlets_j}] in terms of spin observables:
\begin{equation}
\langle {\cal B} \rangle = \langle \sum_{ab} \delta \hat J_a A_{ab} \delta \hat J_b \rangle - \langle \sum_i  \sum_{ab} \hat s^{(i)}_a \tilde{A}_{ab} \hat s^{(i)}_b \rangle
\end{equation}
where $\delta \hat J_a = \hat J_a - \langle \hat J_a \rangle$, $\hat J_a = \sum_{i=1}^N \hat s_a^{(i)}$, and $\hat s_a^{(i)} = \hat S_x^{(i)} \cos \theta_a + \hat S_y^{(i)} \sin \theta_a$. Using $A_{ab}=A_{ba}$, $\tilde{A}_{ab}=\tilde{A}_{ba}$ and elementary algebra, we obtain:
\begin{eqnarray}
	\langle \sum_{ab} \delta \hat J_a A_{ab} \delta \hat J_b \rangle = {\rm Var}(\hat J_x) \sum_{ab} \cos\theta_a A_{ab} \cos\theta_b \nonumber \\
	 + {\rm Var}(\hat J_y) \sum_{ab} \sin\theta_a A_{ab} \sin\theta_b \nonumber \\
	 + (1/2)\langle \{\delta \hat J_x, \delta \hat J_y\}\rangle \sum_{ab} A_{ab} \sin(\theta_a + \theta_b) 
\label{eq_Qdata_singletsBI_1}
\end{eqnarray}
where $\{\hat x, \hat y\} = \hat x \hat y + \hat y \hat x$, and:
\begin{eqnarray}
	\langle \sum_i  \sum_{ab} \hat s^{(i)}_a \tilde{A}_{ab} \hat s^{(i)}_b \rangle = \langle [ \hat S_x^{(i)}]^2\rangle \sum_{ab} \cos\theta_a \tilde{A}_{ab} \cos\theta_b \nonumber \\
	 + \langle [ \hat S_y^{(i)}]^2\rangle \sum_{ab} \sin\theta_a \tilde{A}_{ab} \sin\theta_b \nonumber \\
	 + (1/2)\langle \{\hat S_x^{(i)}, \hat S_y^{(i)}\}\rangle \sum_{ab} \tilde{A}_{ab} \sin(\theta_a + \theta_b) 
\label{eq_Qdata_singletsBI_2}
\end{eqnarray}
We derive explicitly the corresponding witnesses for two families of Bell's inequalities in the next Section. For the sake of illustration, on Fig.~\ref{fig_zoo}(e) we plot the witness condition for the Bell's inequalities we found for $j\in \{1/2,1,3/2,2\}$, with angles $\{\theta_a\}=(-t,0,t)$ [that is, along the diagonal of panels (a-d) of the same Fig.~\ref{fig_zoo}]. To realize this plot and derive a simple condition involving only the collective spin fluctuations ${\rm Var}(\hat J_x)$, we have further assumed $SU(2)$ invariance of the state. With this assumption, we find:
\begin{equation}
\langle {\cal B} \rangle_{SU(2)} =  \langle {\cal B} \rangle_{\rm singlet} +{\rm Var}(\hat J_x) \sum_{ab} e^{-i\theta_a} A_{ab} e^{i\theta_b}
\end{equation}
where $\langle {\cal B} \rangle_{\rm singlet}$ is given by Eq.~\eqref{eq_BQ_singlets_1}. The witness condition is $\langle {\cal B} \rangle_{SU(2)} < -NE_{\rm max}(\tilde{A})$ with $E_{\rm max}(\tilde{A})$ given by Eq.~\eqref{eq_Emax_singlets}. Explicitly, the witness condition for $SU(2)$-invariant states is:
\begin{equation}
	\frac{{\rm Var}(\hat J_x)}{N} < \frac{\min_{\{s_a\}} \sum_{ab}\left[\frac{j(j+1)}{3}e^{i(\theta_b-\theta_a)} - s_a s_b\right]\tilde{A}_{ab}}{\sum_{ab} e^{-i\theta_a} A_{ab} e^{i\theta_b}} ~,
	\label{eq_witness_SU2}
\end{equation}
where the $\min$ is over $\{s_a\}\in \{-j, \dots, j\}^k$. This upper bound is plotted on Fig.~\ref{fig_zoo}(e).
}

\subsubsection{A Bell's inequality for half-integer spin singlets.}
The Bell's inequality presented in Eq.~\eqref{eq_BI_singlets_spin_half}, and valid for $j=1/2$, can be extended to arbitrary spins. Here, we focus on the simplest extension with $k=3$ measurement settings, which corresponds to the region labelled `1' on panels (a) and (c) of Fig.~\ref{fig_zoo}. The coefficients we found are:
\begin{subequations}
	\label{eq_Atilde_h2_1}
	\begin{align}	
	&\tilde{A} = \begin{pmatrix}
		-1 & 1 & -1 \\
		1 & -1 & 1 \\
		-1 & 1 & -1
	\end{pmatrix} \label{eq_Atilde_1} \\
	&h^{(2)} = (3, 3, 3) \label{eq_h2_1}~.
	\end{align}
\end{subequations}
We notice that for arbitrary complex numbers $x_0, x_1, x_2$, we have: $\sum_{ab} x_a^* \tilde{A}_{ab} x_b = -|x_0 - x_1 + x_2|^2 \le 0$. For integer spins, choosing $s_0=s_1=s_2=0$ gives $E_{\rm max}=0$, so that the classical bound cannot be violated by measuring spin singlets [Eq.~\eqref{eq_BQ_singlets_1}]. In contrast, for half-integer spins, the minimal value of $|s_0 - s_1 + s_2|$ is $1/2$, so that $E_{\rm max} = -1/4$. {Instead, for singlets, choosing the measurement angles $\{\theta_a\}=(\pi / 3,0,-\pi/3)$ (which is optimal), we find $\sum_{ab} e^{-i\theta_a} \tilde{A}_{ab} e^{i\theta_b} = 0$. We will now derive a witness condition for this optimal choice of measurements, valid with no assumption about the quantum state. From Eq.~\eqref{eq_Qdata_singletsBI_2}, we find that $\langle \sum_i  \sum_{ab} \hat s^{(i)}_a \tilde{A}_{ab} \hat s^{(i)}_b \rangle=0$. From Eq.~\eqref{eq_Qdata_singletsBI_1}, recalling that $A_{ab}=\tilde{A}_{ab}+\delta_{ab}h_a^{(2)}$, we find $\langle {\cal B} \rangle = (9/2)[{\rm Var}(\hat J_x) + {\rm Var}(\hat J_y)$]. Therefore, violation of the Bell's inequality Eq.~\eqref{eq_BI_SU2_generic}, with coefficients given by Eq.~\eqref{eq_Atilde_h2_1}, and whose classical bound is $B_c=N/4$, is possible whenever:
\begin{equation}
	{\rm Var}(\hat J_x) + {\rm Var}(\hat J_y) < \frac{N}{18} ~. \label{eq_criterion_singlets_1}
\end{equation}
In order to reach this conclusion, we only assumed that a collection of $N$ spin-$j$ particless, with $j$ a half-integer, is measured along well-calibrated axes $x$ and $y$. Maximal violation is obtained for perfect singlets which satisfy ${\rm Var}(\hat J_x)={\rm Var}(\hat J_y)=0$. This generalizes a result of ref.~\cite{frerotR2020} to arbitrary half-integer spins. On Fig.~\ref{fig_zoo}(e) where $SU(2)$ invariance is further assumed (so that ${\rm Var}(\hat J_x) = {\rm Var}(\hat J_y)$), this condition corresponds to the maximum at $t=\pi/3$ for $j\in \{1/2, 3/2\}$. }\\

\subsubsection{A family of Bell's inequalities for arbitrary spin singlets.}
The second family of Bell's inequalities we have characterized is violated {by states sufficiently close to a many-body singlet} for arbitrary $j$, and corresponds to the region labelled `2' on panels (b,c,d) of Fig.~\ref{fig_zoo}. The coefficients of the Bell's inequality are:
\begin{subequations}
	\label{eq_Atilde_h2_2}
	\begin{align}
	&\tilde{A} = 2j\begin{pmatrix}
		-2j & 1 & -2j \\
		1 & 0 & 1 \\
		-2j & 1 & -2j
	\end{pmatrix} \label{eq_Atilde_2}\\
	&h^{(2)} = 8j^2(1, 1, 1) + (1,0,1) \label{eq_h2_2}~.
	\end{align}
\end{subequations}
In this case, for arbitrary complex numbers $x_0, x_1, x_2$, we have: $(2j)^{-1}\sum_{ab} x_a^* \tilde{A}_{ab} x_b = x_1^*(x_0 + x_2) + (x_0^* + x_2^*)[x_1 - 2j(x_0 + x_2)]$. The classical bound is $B_c=0$. Indeed, replacing the complex variables $x_a$ by the variables $s_a \in \{-j, \dots j\}$, we find $E/(2j)^2 := (s_0 + s_2)[s_1/j - (s_0 + s_2)] \le 0$. This quantity is only $\ge 0$ when $s_0 + s_2$ is between $0$ and $s_1/j$. But since $|s_1/j| \le 1$, and since $s_0+s_2$ is always an integer, we conclude that $E_{\rm max}(\tilde A)= 0$. {Concerning the quantum value on spin singlets [Eq.~\eqref{eq_BQ_singlets_1}], the optimal measurement angles [Eq.~\eqref{eq_Qmax_singlets}] are $\{\theta_a\} = (\arccos[1/(4j)], 0, -\arccos[1/(4j)])$, for which we obtain $\sum_{ab} e^{i\theta_a} \tilde{A}_{ab} e^{i\theta_b} = 1$. From Eq.~\eqref{eq_Qdata_singletsBI_2}, we find that $\langle \sum_i  \sum_{ab} \hat s^{(i)}_a \tilde{A}_{ab} \hat s^{(i)}_b \rangle=\sum_{i=1}^N \langle [\hat S_x^{(i)}]^2 \rangle$. We then compute [Eq.~\eqref{eq_Qdata_singletsBI_1}] $\sum_{ab}\sin\theta_a A_{ab}\sin\theta_b=1+16j^2-1/(8j^2)$, $\sum_{ab}\cos\theta_a A_{ab}\cos\theta_b=2+8j^2+1/(8j^2)$, and $\sum_{ab}\sin\theta_a A_{ab}\cos\theta_b=0$, which leads us to the witness condition:
\begin{eqnarray}
	&{\rm Var}(\hat J_x)\left(2 + 8j^2 + \frac{1}{8j^2}\right) + 
	{\rm Var}(\hat J_y)\left(1 + 16j^2 - \frac{1}{8j^2}\right) \nonumber \\
	&-\sum_{i=1}^N \langle [\hat S_x^{(i)}]^2 \rangle < 0
	~. \label{eq_criterion_singlets_2}
\end{eqnarray}
In contrast to Eq.~\eqref{eq_criterion_singlets_1}, this condition is a Bell-correlation witness for arbitrary $j$, and only assumes correct calibration of the measurements. Notice that for $j=1/2$ (for which $[\hat S_x^{(i)}]^2=1/4$), this condition is the same as Eq.~\eqref{eq_criterion_singlets_1}. A simplified witness condition can be obtained by further assuming $SU(2)$ invariance:
\begin{equation}
	\frac{{\rm Var}(\hat J_x)}{N} < \frac{j(j+1)}{9(1+8j^2)} ~.
\end{equation}
  On Fig.~\ref{fig_zoo}(e), this condition corresponds to the right-most maximum at $t=\arccos[1/(4j)]$. \\
}
\noindent\textbf{Further improvement.}
The Bell's inequalities reported here are the simplest ones discovered via our data-driven method, involving only $k=3$ co-planar spin measurements. Adding extra measurements, possibly genuine $SU(d)$ measurements, can only lead to more robust Bell's inequalities, and looser witness conditions than Eqs.~\eqref{eq_criterion_singlets_1} and \eqref{eq_criterion_singlets_2}. We leave this exploration open to future studies, for which our algorithm \cite{code_guillem} represents a natural starting point. 

\subsection{Spin-squeezed states}
\label{sec_squeezing_spin_j}
\noindent\textbf{A Bell's inequality for arbitrary-spin squeezed states with two settings.}
Similarly to the measurement scenario leading to the violation of Eq.~\eqref{eq_BI_Tura} for spin-$1/2$ squeezed states, we consider a situation where two projective spin measurements, $\hat s_{0/1}^{(i)} =\hat S_x^{(i)} \cos \theta  \pm \hat S_y^{(i)} \sin \theta$, are locally performed on a collection of $N$ spin-$j$ particles. Using data from a spin-$j$ squeezed state, we find the following generalization of Eq.~\eqref{eq_BI_Tura}:
\begin{eqnarray}
	\langle {\cal B} \rangle &=& \tilde{C}_{00} + \tilde{C}_{11} - \tilde{C}_{01} - \tilde{C}_{10} + \nonumber \\ 
		&& 2M_0^{(2)} + 2M_1^{(2)} - M_0 - M_1 \ge 0 ~.
		\label{eq_BI_Tura_spin_j}
\end{eqnarray}
For $j=1/2$, we have $M_a^{(2)} = N/4$, and therefore recover Eq.~\eqref{eq_BI_Tura}. More generally, the classical bound $\ge 0$ is obtained by writing ${\cal B}_{\rm LV} = (\delta S_0 - \delta S_1)^2 - \sum_{i=1}^N [(s_0 - s_1)^2 -	2s_0^2 - 2s_1^2 + s_0 + s_1]^{(i)}	\ge -\max_{s_0, s_1} E(s_0, s_1)$,
where $E(s_0, s_1) = (s_0 - s_1)^2 -	2s_0^2 - 2s_1^2 + s_0 + s_1 = -(s_0 + s_1)(s_0 + s_1 - 1)$. Since $s_{0/1}$ are the outcomes of spin-$j$ measurements, they are either both integers, either both half-integers. This implies that $s_0 + s_1$ is always an integer. Since $E(x) = -x(x-1) \le 0$ for all integers $x$, we conclude that $\langle {\cal B} \rangle \ge 0$ for LV models. On the other hand, for the measurement setting we consider, we have the quantum value:
\begin{equation}
	\langle {\cal B} \rangle = 4 {\rm Var}(\hat J_y)\sin^2 \theta - 2\langle \hat J_x \rangle\cos \theta  + 4 \cos^2 \theta \sum_{i=1}^N\langle [\hat S_x^{(i)}]^2 \rangle
\end{equation}
We introduce the notation $s_x^2 = N^{-1} \sum_{i=1}^N\langle [\hat S_x^{(i)}]^2 \rangle$. The optimal measurement angle $\theta$, leading to the minimal value of $\langle {\cal B} \rangle$, is s.t. $\cos \theta = \langle \hat J_x \rangle / [4Ns_x^2 - 4{\rm Var}(\hat J_y)]$ (if this is $\le 1$), for which we obtain:
\begin{equation}
	\langle {\cal B} \rangle = 4 {\rm Var}(\hat J_y) - \frac{\langle \hat J_x^2 \rangle}{4[Ns_x^2 - {\rm Var}(\hat J_y)]} ~.
	\label{eq_witness_squeezing_spin_j}
\end{equation} 
Violation is detected whenever $\langle {\cal B} \rangle < 0$. This generalizes the results for $j=1/2$ \cite{schmiedetal2016,piga_bell_2019} to arbitrary spins. In Appendix \ref{sec_app_PIBI_spin_squeezed}, we present another Bell's inequality, for which violation has been found with $j\le 1$. Clearly, adding extra measurements ($k \ge 3$) could only lead to more robust Bell's inequalities (see Sec.~\ref{sec_PIBI_squeezed} for the case $j=1/2$): we leave this exploration open to future works. \\

\section{Experimental implementation}
\label{sec_experimental}
In this work, we have introduced a new methodology to learn, from experimental data themselves, the best device-independent entanglement criterion that the data allow one to construct -- in the form of a Bell's inequality whose coefficients are inferred via a data-driven algorithm. We have demonstrated the effectiveness of this new approach by using, as input to our algorithm, either actual experimental data \cite{schmiedetal2016}, or data which could be obtained by collecting the appropriate two-body correlations on realistic many-body quantum states. The new Bell's inequalities presented in this work include and surpass all robust permutationally-invariant Bell's inequalities reported so far in the literature. Therefore, these Bell's inequalities could already be useful for entanglement certification in existing or near-term quantum devices, if the relevant states are prepared, and if the appropriate measurements are performed (see below). But most importantly, the data-driven nature of our approach makes it especially suitable to explore a virtually-infinite variety of experimental data, potentially unveiling new and unexpected Bell's inequalities. It is indeed not unrealistic to anticipate that present-day quantum simulators and computers are processing quantum-entangled states, while the experimentalists are not able to prove it simply because they lack entanglement criteria tailored to their experimental data. To facilitate this exploration by other researchers, we have released a pedagogical open-access version of the code used throughout this paper \cite{code_guillem}. Furthermore, in this section, we present an (incomplete) list of experimental platforms able to produce suitable data, allowing one to potentially certify the preparation of many-body entangled state with the data-driven method presented in this work. { The present section does not contain new results; rather, it consists of a guide towards the existing literature relevant to the experimental implementation of device-independent entanglement certification. Such implementation} requires answering two questions:
   \begin{itemize}
\item A) Can one collect the data sets used as inputs to our data-driven algorithm?
\item B) Can one prepare quantum many-body states manifesting Bell's non-locality?
 \end {itemize}

The measurement question can take two conceptually-different forms: A1) the one- and two-body correlations forming the data set [Eqs.~\eqref{eq_def_C}, \eqref{eq_def_Ct} and \eqref{Qdata_Ma2}] are measured individually, which requires individual addressing of the subsystems; A2) these data are inferred from the fluctuations of collective observables. In the first case (A1), one realizes a situation conceptually close to a proper Bell test, even though avoiding e.g. the locality loophole might be very challenging. In contrast, in the second case (A2), one does not realize a Bell test, but rather demonstrates the ability to prepare many-body entangled states which would yield violation of the reconstructed Bell's inequalities, if such a Bell test could be implemented. Clearly, (A2) is less demanding and requires only access to collective variables (as realized in \cite{schmiedetal2016,engelsen_bell_2017}), which are sums of spins or pseudo-spins of individual components of the system. These individual components could be atoms of spin $j=1/2, 1, 3/2, \dots$, or atoms, ions and superconducting qubits realizing effective few-level systems. First- and second-moments of the collective-spin components must be measured. { For spin-$j$ particles, witness conditions such as those we derived [Eq.~\eqref{eq_criterion_singlets_2} and \eqref{eq_witness_squeezing_spin_j}] may involve quantities of the form $\hat {\cal O}=\sum_{i=1}^N f(\hat S_{\bf n}^{(i)})$, where $\hat S_{\bf n}^{(i)}$ is a spin observable along an arbitrary direction ${\bf n}$, and $f(x)$ may be an arbitrary function. Such observables can be measured via collective measurements in atomic systems, e.g. after Stern-Gerlach splitting of the magnetic sublevels before imaging \cite{nayloretal2016}. Indeed, we may express $\hat {\cal O} = \sum_{s=-j}^j \hat P_s^{({\bf n})} f(s)$, where $\hat P_s^{({\bf n})}$ is the number of atoms detected with spin $s$ after the Stern-Gerlach splitting with a magnetic field gradient along ${\bf n}$.}  

Concerning the preparation (B), the goal is to generate strongly entangled many-body states, such as squeezed states or spin singlets. %which appear as ground states of Hamiltonians with long-range interactions, such as, for instance, the ones corresponding to Lipkin-Meshkov-Glick models (see \cite{Bao2020} and references therein). 
Notice that while spin singlets and spin squeezed states are paradigmatic examples of many-body entangled states, on which we focused in this work to demonstrate the effectiveness and flexibility of our new method, other classes of states are potentially interesting; for instance, Dicke states are clear candidates \cite{Lucke773,Zou6381} while the potentialities of $j\ge 1$ ensembles subjected to more general $SU(d)$ measurements, are virtually unlimited. In this section, we present an incomplete list of platforms for which both A and B questions can be answered positively. { We then discuss the experimental effort, in terms of the number of repetitions of the preparation-measurement procedure, in order to establish the violation of permutationally-invariant Bell's inequalities.}\\

\subsection{Experimental platforms}
\label{sec_exp_platforms}
 \noindent   {\it Atomic ensembles.} These are clouds of (not-necessarily cold) atoms with spin. A) The total spin components can be measured employing quantum Faraday effect, i.e. looking at the polarisation rotation of the light passing through the atomic cloud \cite{Klemens}. This method is also frequently termed as spin polarisation spectroscopy (SPS). In principle, one has access here to the full quantum statistics of the total spin components. Using standing driving fields, one can also detect spatial Fourier components of the collective spin \cite{MLPolzik}. % Repeated measurements give access to temporal correlations \cite{MLOriol}. 
 B) Atomic ensembles are particularly suitable to achieve strong squeezing of the atomic spin. Using quantum feedback, spin singlet states have also been prepared \cite{MorganToth,MorganToth1,MorganExp}. Combining feedback with the ideas of Ref. \cite{MLPolzik}, practically arbitrary spin-spin correlations could be generated \cite{HaukeMorgan}. Using a one-dimensional atom-light interface, quantum spin noise limit can be achieved \cite{Polzik2018}. Spin-squeezed states and even Bell's non-locality have been achieved for macroscopic ensembles, using optical-cavity-assisted measurements \cite{engelsen_bell_2017}. \\

   \noindent {\it Ultracold spinor Bose-Einstein condensates.} A) The techniques developed for atomic ensembles can be applied to Bose-Einstein condensates \cite{Eckert}. In ultracold trapped spinor gases, the principal non-linear mechanism leading to squeezing (among other interesting entangled states) corresponds to spin-changing collisions \cite{Sengstock}. Here, all spin components and their fluctuations can be measured. Beyond spin components, e.g. the nematic tensor for spin-1 condensates can be measured, including in a spatially-resolved way \cite{kunkeletal2019}. %One can even measure quantities relevant to metrology, such as Fisher information, that can serve as entanglement witnesses \cite{Markus}. Both atomic ensembles and spinor condensates are especially suitable for magnetometry (cf. \cite{MitchellRMP,MorganNJP}. 
   B) The possibility to violate the many-body Bell's inequalities of Ref.~\cite{turaetal2014} were in fact first confirmed in twin-modes squeezed states in spinor condensates \cite{schmiedetal2016}, and these systems allow one to generate non-classical states going beyond squeezing (for a review, see \cite{pezzeetal2016}). Entanglement between spatially separated condensates, and even Einstein-Podolsky-Rosen steering, can be generated \cite{Fadel,kunkeletal2018,Klempt}. More recently, high-spin cold-atoms ensembles, which display dipolar magnetic interactions, have been generated \cite{batailleetal2020,trautmannetal2018,tanzi_supersolid_2019,bottcheretal2020}, which appear as ideal playgrounds to explore Bell's inequalities tailored to arbitrary-spin systems, as established in the present work. \\

 \noindent{\it Ultracold atoms in optical lattices.} A) Ultracold atoms in optical lattices provide one of the best platforms for quantum simulations \cite{LSA2017book}. All the methods mentioned above can be carried over to atoms in optical lattices. SPS has emerged as a promising technique
for detecting quantum phases in lattice gases via the coherent mapping of spin-correlations onto scattered light, realizing quantum non-demolition measurements. In particular, spatially-resolved SPS that employs standing wave laser configurations \cite{MLPolzik} allows for a direct probing of magnetic
structure factors and order parameters \cite{roscilde_quantum_2009,de_chiara_probing_2011,weitenberg_coherent_2011,meineke_interferometric_2012}. Moreover, quantum gas microscopes, which are able to resolve individual atoms located in single lattice sites, have been developed \cite{Greiner,Kuhr,Ott}. These techniques allow for a direct inspection into the spatial structure of entanglement within the system \cite{fukuharaetal2015}, and direct violation of the Bell's inequalities (A1) could be envisioned. B) These systems may lead to a very large variety of strongly-correlated many-body states. In particular, spin singlets, which are ground states of quantum antiferromagnets according to theorems by Mattis \cite{Auerbach}, are expected to emerge at low-energy in quantum simulators of the Fermi-Hubbard model \cite{TARRUELL2018365}. A review of potentially achievable correlations can be found in Ref.~\cite{Chiara}. %Perhaps the most interesting recent example deals with realisation of lattice gauge thoery with a bosonic Hubbard model \cite{HaukeWan}. 
High-spin atomic ensembles in optical lattices \cite{gabardosetal2020,patscheideretal2020} clearly represent very promising systems to investigate novel classes of entangled many-body states, especially concerning Bell's inequalities involving many outcomes.\\

\noindent{\it Trapped ions.} A) Trapped ions represent a very versatile platform, and one of the most promising candidates for quantum computing. Small systems of ions allow for full quantum tomography of the density matrix, from which the statistics of all observables can be recovered. This includes in particular spin-spin correlations (see for instance \cite{Brydges260}), but also e.g. third order correlations which have been used for the detection of genuine three-body entanglement in a system of few ions. Clearly, trapped ions are a platform of choice to directly probe the spatial structure of quantum correlations, and direct violation of the Bell's inequalities could be achieved (A1). B) Few-ion systems can be used for the generation of a very wide variety of entangled states on demand: recent examples include ground states of lattice gauge theory models, \cite{LGT2020,LGTreview, Nature-blatt}, and dynamically-generated entanglement \cite{Monroe}, among others.\\

 \noindent{\it Significant others.} These include, but are not limited to, atoms in nano-structures \cite{Darrick}, Rydberg atoms \cite{Lukin}, and a large number of
 condensed matter systems, ranging from circuit QED, through quantum dots, to superconducting Josephson junctions \cite{Martinis}. All of these systems could potentially prepare and detect the entangled states suitable to violate the Bell's inequalities investigated in this work -- and most importantly, generate correlation patterns from which data-driven methods such as ours could reveal novel Bell's inequalities.\\ 
 
{
\subsection{Measurement effort} 
\label{sec_experimental_noise}
In order to implement the data-driven method presented in this paper, one needs to estimate the data $M_{a}$ and $\tilde{C}_{ab}$ [Eq.~\eqref{eq_def_Ct}] (the generalization to include also terms such as $M_a^{(2)}$ [Eq.~\eqref{Qdata_Ma2}] for measurements with $d>2$ outcomes is straightforward). In order to evaluate them directly to realize a device-independent entanglement test (i.e. without inferring them via collective measurements to estimate Bell correlation witnesses), one needs to have the experimental capability to individually address each subsystem, and to choose independently the measurement setting $a \in \{0, \dots, k-1\}$ on each of them. The following procedure may be repeated $R$ times.
\begin{enumerate}
	\item Choose randomly and independently a measurement setting $a_i(r)$ on each subsystem $i \in \{1, \dots, N\}$, with a uniform probability over $\{1, \dots, k\}$. ($r \in \{1, \dots, R\}$ labels the $r$-th measurement run);
	\item Perform the corresponding measurement, collecting the string of outcomes ${\bf s}(r)=(s_1(r), \dots, s_N(r))$.
\end{enumerate}
We denote as $R_a^{(i)}$ the number of times the setting $a$ has been implemented on subsystem $i$, and $R_{ab}^{(ij)}$ the number of times the pair of settings $(a,b)$ has been implemented on the pair of subsystems $(i,j)$, i.e.:
\begin{subequations}
\begin{align}
	&R_a^{(i)} = \sum_{r=1}^R \delta_{a,a_i(r)} \\
	&R_{ab}^{(ij)} = \sum_{r=1}^R \delta_{a,a_i(r)} \delta_{b,a_j(r)} ~.
\end{align}
\end{subequations}
On average, for each subsystem, each setting $a$ is implemented $R/k$ times; and for each pair of subsystems, each pair of settings $(a, b)$ is implemented $R/k^2$ times. The data are then obtained as:
\begin{subequations}
\begin{align}
	&M_a^{(\rm exp)} =  \sum_{i=1}^N \frac{1}{R_a^{(i)}} \sum_{r=1}^R \delta_{a,a_i(r)} s_i(r) \\
	&C_{ab}^{(\rm exp)} =  \sum_{i \neq j} \frac{1}{R_{ab}^{(ij)}} \sum_{r=1}^R \delta_{a,a_i(r)}\delta_{b,b_j(r)} s_i(r)s_j(r) \\
	&\tilde{C}_{ab}^{(\rm exp)} = C_{ab}^{(\rm exp)} - M_a^{(\rm exp)} M_b^{(\rm exp)} ~.
\end{align}
\end{subequations}
The collective quantity $M_a$ is typically scaling as $O(N)$ with fluctuations of order $O(\sqrt{N})$ (this holds whenever there is a finite correlation length in the system). On the other hand, the collective quantity $C_{ab}$ scales as $O(N^2)$ with fluctuations of order $O(N)$. The quantity $\tilde{C}_{ab}$ is instead scaling as $O(N)$, but its fluctuations, stemming from the fluctuations of $C_{ab}$ and $M_a M_b$ which are both of order $O(N)$, are also of order $O(N)$. Therefore, the error on $M_a$ and $\tilde{C}_{ab}$ due to finite statistics scale as:
\begin{subequations}
\begin{align}
	&|M_a^{(\rm exp)} - M_a| =  O\left(\sqrt{\frac{N}{R/k}}\right) \\
	&|\tilde{C}_{ab}^{(\rm exp)}- \tilde{C}_{ab}| = O\left(\frac{N}{\sqrt{R/k^2}}\right)
\end{align}
\end{subequations}
and the relative errors scale according to $O\left(\sqrt{\frac{k}{RN}}\right)$ and $O\left(\sqrt{\frac{k^2}{R}}\right)$, respectively. The most demanding estimation is for the two-body correlations contained in $\tilde{C}_{ab}$. Notice that, as a consequence of the fact that the data involve only extensive quantities, the number $R$ of measurement runs required to reach a given relative precision of $\epsilon$ scales as $R \sim k^2/\epsilon^2$, and therefore does not scale with the system size. Notice also that if the goal is not to collect the data to be used as input of our data-driven algorithm, but instead to evaluate the violation of a given permutationally-invariant Bell's inequality, such as those presented in this work, it might be more efficient to select the measurement settings with probabilities depending on the coefficients of the Bell's inequality in question. One then needs to estimate the error on $\sum_{ab}A_{ab} \tilde{C}_{ab} + \sum_a h_a C_a - B_{\rm c}$ (with $B_{\rm c}$ the classical bound), which should be significantly negative for the certification to be conclusive (under the gaussian-statistics assumption, see below). \\

\noindent \textit{Improvement due to the non-linear nature of the Bell's inequalities.}
As already noticed in Section \ref{sec_PIBI_squeezed}, the fact that our Bell's inequalities involve the (non-linear) $\tilde{C}_{ab}$ quantities leads to significantly tighter results than the Bell's inequalities involving $C_{ab}$ (see in particular Fig.~\ref{fig_tilde_vs_nontilde}), especially for $N$ large. Indeed, $C_{ab}$ is typically of order $O(N^2)$, while $M_a, \tilde{C}_{ab}=O(N)$, together with the classical bound which is also $O(N)$ [see Eq.~\eqref{eq_BI_Ct} or Eqs.~\eqref{eq_BI_singlets_spin_half} and \eqref{eq_BI_Tura} for explicit examples]. Therefore, any systematic error will lead to an error of $O(N^2)$ on $C_{ab}$, making more challenging in practice the detection of Bell non-locality for large $N$ (see, however, refs.~\cite{schmiedetal2016,engelsen_bell_2017}). Instead, given that all terms in our Bell's inequalities are extensive, a systematic error will lead to $O(N)$ deviations, which does not represent an obstruction to scalable Bell tests.\\

\noindent \textit{Relaxing the gaussian-statistics assumption.}
If the implicit gaussian-statistics assumption leading to the scaling $R \sim k^2/\epsilon^2$ is to be relaxed, more elaborate finite-statistics analysis must be carried on, typically using tail bounds on the distribution of outcomes (see e.g. refs.~\cite{zhang_asymptotically_2011,elkouss_nearly_2016,klieschR2021}). Similarly, if a Bell correlation witness, based on collective measurements, is to be evaluated, from which the ability of the prepared state to violate a Bell's inequality is to be assessed, special care in the data analysis must be taken if the gaussian-statistics assumption is relaxed \cite{wagner_bell_2017}, leading to an overhead in terms of measurement effort.
}
 \subsection{Summary of the concrete implementation of our method.} In summary, detecting entanglement via our data-driven method proceeds in four steps:
 \begin{enumerate}
 \item Define a partition of the multipartite system into $N$ subsystems, and select several (incompatible) local quantum observables ${\hat s}_a^{(i)}$ for $i \in \{1, \dots N\}$, whose outcome are denoted $s_a^{(i)}$;
 \item Collect one-body terms $\sum_{i=1}^N \langle s_a^{(i)} \rangle$, and two-body terms $\sum_{i \neq j} \langle s_a^{(i)} s_b^{(j)} \rangle $, either by measuring individually the subsystems, or by inferring such data via collective measurements;
 \item Use these data as input to our algorithm to potentially find a violated Bell's inequality \cite{code_guillem}. If no violation is found, one could modify the measurements chosen at step (1);
 \item Analytically analyze the Bell's inequality inferred from the data, to understand the essential features leading to entanglement detection.
 \end{enumerate}
 
\section{Conclusions}
\label{sec_conclusions}
We have presented a new data-driven method to detect multipartite entanglement in quantum simulators and computers. We devised an algorithm (Section \ref{sec_convex_opt_algo_d2}) which constructs {a violated non-linear Bell's inequality from one- and two-body correlations averaged over all permutations of the subsystems. Our approach is applicable to any number of measurement outcomes. In order to do so, we have expressed the two-body coefficients of the Bell's inequality as a positive semidefinite matrix, whose optimization allows for a systematic exploration of all potentially violated Bell's inequalities of this form.} As an illustration of the potentialities of this new approach to entanglement detection, we could improve over previously-known many-body Bell's inequalities violated by $j=1/2$ spin-squeezed \cite{turaetal2014,schmiedetal2016,engelsen_bell_2017,wagner_bell_2017} and spin singlet states \cite{frerotR2020} in the thermodynamic limit (Section \ref{sec_d2}). In addition, we could extend these results to similar states for arbitrary $j>1/2$ individual spins, by considering Bell scenarios with arbitrarily-many outcomes (Section \ref{sec_many_outcomes}) -- to our knowledge, this represents the first example of such families of many-body Bell's inequalities. {As our (non-linear) Bell's inequalities involve only zero-momentum fluctuations, sufficient conditions on many-body quantum states for their violation could be established, in the form of Bell-correlation witnesses -- involving first moments and variances of collective observables. Such witnesses can be measured in state-of-the-art cold-atoms systems with only global measurements (Section \ref{sec_experimental}). Importantly, the non-linear nature of the Bell's inequalities reconstructed by our method offers a fundamental scaling improvement over the linear Bell's inequalities which have been considered so far.} 
Due to its very flexible nature and a very small computational cost (independent of the system size, and exponential in the number of measurement outcomes), our data-driven approach opens the way to the systematic exploration of permutationally-invariant Bell's inequalilites in many-qudits systems -- as a matter of fact, the Bell's inequalities presented in the paper represent only a fraction of all those discovered with our approach \cite{code_guillem}, already for the simple classes of spin-squeezed and spin-singlet states. We anticipate that exploring other many-body entangled states, for instance considering Dicke states, or going beyond spin measurements to consider genuine $SU(d)$ measurement \cite{kunkeletal2019}, either theoretically, or directly from experimental data, will lead to the discovery of yet many other and -- by construction -- useful many-body Bell's inequalities. 
Finally, we would like to point out that our algorithm searches for Bell's inequalities whose two-body coefficients form a positive semi-definite matrix. While all robust permutationally-invariant Bell's inequality reported in the literature satisfy this condition, it is worth investigating its limitations.

\acknowledgments{We acknowledge the Spanish Ministry MINECO (National Plan 15 Grant: FISICATEAMO No. FIS2016-79508-P, SEVERO OCHOA No. SEV-2015-0522, FPI, FIS2020-TRANQI and Severo Ochoa CEX2019-000910-S), European Social Fund, Fundació Cellex, Fundació Mir-Puig, Generalitat de Catalunya (AGAUR Grant No. 2017 SGR 1341, AGAUR SGR 1381, CERCA program, QuantumCAT\_U16-011424, co-funded by ERDF Operational Program of Catalonia 2014-2020), MINECO-EU QUANTERA MAQS (funded by The State Research Agency (AEI) PCI2019-111828-2 / 10.13039/501100011033), and the National Science Centre, Poland-Symfonia Grant No. 2016/20/W/ST4/00314. IF acknowledges support from the Mir-Puig and Cellex fundations through an ICFO-MPQ Postdoctoral Fellowship.}

\appendix

\section{Convex optimization algorithm based on the averaged pair probability distribution}
\label{sec_app_algo}
In this Section, we present a general formulation of the convex optimization algorithm to find a Bell's inequality violated by the pair probability distribution averaged over all permutations of the subsystems [Eq.~\eqref{eq_Pbar}]:
\begin{equation}
	\bar{P}(s, t|a, b) = \frac{1}{N(N-1)} \sum_{i \neq j} P^{(ij)}(s, t|a, b) ~.
\end{equation}
The probability distribution $\bar{P}$ is the central object reconstructed from experimental observations. $a,b=0, \dots k-1$ label the measurement settings, while $s,t=1, \dots d$ label the measurement outcomes.  Assuming that a LV model $P_{\rm LV}({\bm \sigma})$ exists, which returns ${P}^{(ij)}$ as a marginal, with ${\bm \sigma} = \{s_a^{(i)}\}$ a collection of classical variables representing the measurement outcomes, we have [Eq.~\eqref{eq_Ppair_marginal_PLV}]:
\begin{equation}
	P^{(ij)}(s, t|a, b) = \sum_{\bm \sigma}P_{\rm LV}({\bm \sigma}) \delta_{s_a^{(i)}, s} \delta_{s_b^{(j)}, t} ~.
\end{equation}
We may then decompose $\bar{P}$ as:
\begin{equation}
	\bar{P} = \frac{1}{N(N-1)} \left[\sum_{i, j} P^{(ij)} - \sum_{i} P^{(ii)}\right] ~.
\end{equation}
Even though $P^{(ii)}(s,t|a,b)$ is not observable (since it would require measuring simultaneously the settings $a$ and $b$ on the same subsystem $i$, and in general they correspond to incompatible quantum observables), it exists at the level of the LV model. The key point is then that if we define $Q_{(s,a), (t,b)} = \sum_{i, j} P^{(ij)}(s,t|a,b)$, then $Q$ [as a $(kd) \times (kd)$ matrix] is PSD. Indeed, considering a $kd$-component vector $f(s|a)$, we have:
\begin{eqnarray}
	f^T Q f &=& \sum_{s,t}\sum_{a,b} \sum_{i,j} P^{(ij)}(s,t|a,b) ~f(s|a) ~f(t|b) \nonumber \\
	&=& \sum_{\bm \sigma} P_{\rm LV}({\bm \sigma}) \sum_{i,j} \sum_{a,b} \sum_{s,t} \delta_{s_a^{(i)},s} f(s|a)~ \delta_{s_b^{(j)}, t} f(t|b) \nonumber \\
	&=& \sum_{\bm \sigma} P_{\rm LV}({\bm \sigma}) \sum_{i,j} \sum_{a,b}  f[s_a^{(i)}|a]~ f[s_b^{(j)}|b] \nonumber \\
	&=& \sum_{\bm \sigma} P_{\rm LV}({\bm \sigma}) \left\{\sum_{i} \sum_{a}  f[s_a^{(i)}|a]\right\}^2 \nonumber \\
	& \ge & 0 ~.
\end{eqnarray}
Consequently, for any PSD matrix $M_{(s,a),(t,b)} := M(s,t|a,b)$, we have:
\begin{eqnarray}
	&\sum_{a,b} \sum_{s,t} M(s,t|a,b) \bar{P}(s,t|a,b) \ge \nonumber \\
	& -\frac{1}{N(N-1)} \sum_i \sum_{a,b} \sum_{s,t} M(s,t|a,b) P^{(ii)}(s,t|a,b) \nonumber \\
	& \ge  -\frac{1}{N-1} E_{\rm max}(M) ~,
\end{eqnarray}
where:
\begin{equation}
	E_{\rm max}(M)= \max_{{\bf s} \in \{1, \dots d\}^k} \sum_{a,b} M(s_a, s_b|a,b) ~.
\end{equation}
The optimal PSD matrix $M$ may therefore be found by a convex-optimization program, minimizing the cost function: 
\begin{equation}
	L(M) = \sum_{a,b} \sum_{s,t} M(s,t|a,b) \bar{P}(s,t|a,b) + \frac{1}{N-1} E_{\rm max}(M) 
\end{equation}
over all PSD matrices $M$.

\section{Bell's inequality for spin-squeezed states}
\label{sec_app_PIBI_spin_squeezed}
Exploring $j=1$ spin-squeezed states with $k=2$ spin measurements in the $xy$-plane, at angle $\pm \theta$ with respect to the $x$ axis, we found another violated Bell's inequality similar to the one presented in Section \ref{sec_PIBI_squeezed}:
\begin{eqnarray}
	\langle {\cal B} \rangle = \tilde{C}_{00} + \tilde{C}_{11} - \tilde{C}_{01} - \tilde{C}_{10} + \nonumber \\ 
		M_0^{(2)} + M_1^{(2)} - 2M_0 - 2M_1
		 \\
		=\langle\delta(S_0 - S_1)^2\rangle + 2\sum_{i=1}^N\langle s_0 s_1 - s_0  - s_1 \rangle^{(i)}\\
		\ge -2Nj^2 ~.
\end{eqnarray}
The quantum value is (with $s_a^2 = N^{-1}\sum_{i=1}^N \langle [\hat S_a^{(i)}]^2 \rangle$ for $a=x,y$): 
\begin{equation}
	\langle {\cal B} \rangle = 4{\rm Var}(\hat J_y)\sin^2\theta  -4\langle \hat J_x \rangle \cos \theta + 2N(s_x^2\cos^2 \theta  - s_y^2\sin^2 \theta )
\end{equation}
The optimal angle is s.t. $\cos\theta = \langle \hat J_x \rangle / [Ns_x^2 + Ns_y^2 - 2 {\rm Var}(\hat J_y)]$, for which we have:
\begin{equation}
	\langle {\cal B} \rangle = 2[2{\rm Var}(\hat J_y) - Ns_y^2] - \frac{2\langle \hat J_x \rangle^2}{Ns_x^2 + Ns_y^2 - 2{\rm Var}(\hat J_y)}
\end{equation}
We found violation for squeezed states of $j=1/2$ or $j=1$.

\bibliography{biblio}

\end{document}